\documentclass[onecolumn]{IEEEtran}

\usepackage[latin1]{inputenc}
\usepackage[english]{babel}
\usepackage[margin=1in]{geometry}

\usepackage{blindtext}
\usepackage{amssymb}
\usepackage{hyperref}
\hypersetup{
    colorlinks=true,
    linkcolor=blue,
    filecolor=magenta,      
    urlcolor=cyan,
}
 
\usepackage{multirow}
\usepackage{makecell}
\usepackage{amsthm}
\usepackage[normalem]{ulem}
\usepackage{amsmath}
\usepackage{mathrsfs}
\usepackage{booktabs}
\usepackage{subfigure}
\usepackage{array}
\usepackage[noadjust]{cite}
\usepackage{soul}
\usepackage{mathtools}
\usepackage{relsize}
\usepackage{bm}
\usepackage{tabularx}
\usepackage{algorithm}
\usepackage{algpseudocode}
\usepackage{float}
\newtheorem{theorem}{Theorem}[section]
\newtheorem{corollary}{Corollary}[section]
\newtheorem{lemma}[theorem]{Lemma}
\newtheorem{proposition}[theorem]{Proposition}
\newtheorem{construction}{Construction}

\newtheorem{example}{Example}
\newtheorem{remark}{Remark}[section]

\newcommand\nc\newcommand
\nc{\cA}{\mathcal{A}}\nc{\cB}{\mathcal{B}}\nc{\cC}{\mathcal{C}}\nc{\cD}{\mathcal{D}}
\nc{\cE}{\mathcal{E}}\nc{\cF}{\mathcal{F}}\nc{\cG}{\mathcal{G}}\nc{\cH}{\mathcal{H}}
\nc{\cI}{\mathcal{I}}\nc{\cJ}{\mathcal{J}}\nc{\cK}{\mathcal{K}}\nc{\cL}{\mathcal{L}}
\nc{\cM}{\mathcal{M}}\nc{\cN}{\mathcal{N}}\nc{\cO}{\mathcal{O}}\nc{\cP}{\mathcal{P}}
\nc{\cQ}{\mathcal{Q}}\nc{\cR}{\mathcal{R}}\nc{\cS}{\mathcal{S}}\nc{\cT}{\mathcal{T}}
\nc{\cU}{\mathcal{U}}\nc{\cV}{\mathcal{V}}\nc{\cW}{\mathcal{W}}\nc{\cX}{\mathcal{X}}
\nc{\cY}{\mathcal{Y}}\nc{\cZ}{\mathcal{Z}}

\nc{\bba}{\mathbf{a}}\nc{\bbb}{\mathbf{b}}\nc{\bbc}{\mathbf{c}}\nc{\bbd}{\mathbf{d}}
\nc{\bbe}{\mathbf{e}}\nc{\bbf}{\mathbf{f}}\nc{\bbg}{\mathbf{g}}\nc{\bbh}{\mathbf{h}}
\nc{\bbi}{\mathbf{i}}\nc{\bbj}{\mathbf{j}}\nc{\bbk}{\mathbf{k}}\nc{\bbl}{\mathbf{l}}
\nc{\bbm}{\mathbf{m}}\nc{\bbn}{\mathbf{n}}\nc{\bbo}{\mathbf{o}}\nc{\bbp}{\mathbf{p}}
\nc{\bbq}{\mathbf{q}}\nc{\bbr}{\mathbf{r}}\nc{\bbs}{\mathbf{s}}\nc{\bbt}{\mathbf{t}}
\nc{\bbu}{\mathbf{u}}\nc{\bbv}{\mathbf{v}}\nc{\bbw}{\mathbf{w}}\nc{\bbx}{\mathbf{x}}
\nc{\bby}{\mathbf{y}}\nc{\bbz}{\mathbf{z}}

\nc{\bbA}{\mathbf{A}}\nc{\bbB}{\mathbf{B}}\nc{\bbC}{\mathbf{C}}\nc{\bbD}{\mathbf{D}}
\nc{\bbE}{\mathbf{E}}\nc{\bbF}{\mathbf{F}}\nc{\bbG}{\mathbf{G}}\nc{\bbH}{\mathbf{H}}
\nc{\bbI}{\mathbf{I}}\nc{\bbJ}{\mathbf{J}}\nc{\bbK}{\mathbf{K}}\nc{\bbL}{\mathbf{L}}
\nc{\bbM}{\mathbf{M}}\nc{\bbN}{\mathbf{N}}\nc{\bbO}{\mathbf{O}}\nc{\bbP}{\mathbf{P}}
\nc{\bbQ}{\mathbf{Q}}\nc{\bbR}{\mathbf{R}}\nc{\bbS}{\mathbf{S}}\nc{\bbT}{\mathbf{T}}
\nc{\bbU}{\mathbf{U}}\nc{\bbV}{\mathbf{V}}\nc{\bbW}{\mathbf{W}}\nc{\bfX}{\mathbf{X}}
\nc{\bbY}{\mathbf{Y}}\nc{\bbZ}{\mathbf{Z}}

\nc{\sA}{\mathsf{A}}\nc{\sB}{\mathsf{B}}\nc{\sC}{\mathsf{C}}\nc{\sD}{\mathsf{D}}
\nc{\sE}{\mathsf{E}}\nc{\sF}{\mathsf{F}}\nc{\sG}{\mathsf{G}}\nc{\sH}{\mathsf{H}}
\nc{\sI}{\mathsf{I}}\nc{\sJ}{\mathsf{J}}\nc{\sK}{\mathsf{K}}\nc{\sL}{\mathsf{L}}
\nc{\sM}{\mathsf{M}}\nc{\sN}{\mathsf{N}}\nc{\sO}{\mathsf{O}}\nc{\sP}{\mathsf{P}}
\nc{\sQ}{\mathsf{Q}}\nc{\sR}{\mathsf{R}}\nc{\sS}{\mathsf{S}}\nc{\sT}{\mathsf{T}}
\nc{\sU}{\mathsf{U}}\nc{\sV}{\mathsf{V}}\nc{\sW}{\mathsf{W}}\nc{\sX}{\mathsf{X}}
\nc{\sY}{\mathsf{Y}}\nc{\sZ}{\mathsf{Z}}

\newcommand{\abs}[1]{\left|#1\right|}

\newcommand{\floorenv}[1]{\left\lfloor #1 \right\rfloor}

\nc{\set}[1]{\llbracket #1 \rrbracket}
\newcommand{\Int}[1]{{\left[{#1}\right]}}
\newcommand{\Integers}{{\mathbb{Z}}}
\newcommand{\code}{{\mathcal{C}}}

\newcommand{\Support}{{\mathsf{Supp}}}

\newcommand{\bal}[1]{\begin{align}\label{#1}}
\newcommand{\eal}{\end{align}}
\renewcommand{\le}{\leqslant}
\renewcommand{\leq}{\leqslant}
\renewcommand{\ge}{\geqslant}
\renewcommand{\geq}{\geqslant}

\newcommand{\ms}[1]{\{\{ #1 \}\}}

\theoremstyle{definition}
\newtheorem{definition}{Definition}[section]

\title{New Bounds and Constructions for Variable Packet-Error Coding}
\author{Xiangliang~Kong, Xin~Wang, Ron~M.~Roth, and Itzhak~Tamo%
\thanks{This work was supported in part by the European Research Council (ERC) under Grant 852953 and in part by the Israel Science Foundation under Grant 2369/24.}
\thanks{X. Kong (rongxlkong@gmail.com) and I. Tamo (tamo@tauex.tau.ac.il) are with the School of Electrical \& Computer Engineering, Tel Aviv University, Tel Aviv-Yafo 6997801, Israel. X. Wang (xinw@suda.edu.cn) is with the Department of Mathematics, Soochow University, Suzhou 215005, Jiangsu, China. Ron~M.~Roth (ronny@cs.technion.ac.il) is with the Computer Science Department, Technion, Haifa 3200003, Israel.}
}
\date{today}

\begin{document}

\maketitle

\begin{abstract}
In this paper, we consider the problem of variable packet-error coding, which emerges in network communication scenarios where a source transmits information to a destination through multiple disjoint paths. The objective is to design codes with dynamic error-correcting capabilities that adapt to varying numbers of errors. Specifically, we first provide several bounds on the rate--distortion trade-off for general variable packet-error coding schemes. Then, we present two explicit constructions of variable packet-error coding schemes. The first construction uses higher-order MDS codes and provides a coding scheme that achieves a better rate--distortion trade-off compared to known results for general parameter regimes. The second construction is based on a variant of the repetition code and yields a coding scheme with an optimal rate--distortion trade-off, with respect to our bound, for certain parameter regimes.
\end{abstract}

\section{Introduction}

\noindent

Since the publication of \cite{ACLY20}, network coding theory has emerged as one of the most influential and dynamic branches of coding theory and information theory. Over the years, researchers from different fields have shown significant interests in network coding theory due to its broad range of applications in practical scenarios. Within this field, network error-correcting (NEC) codes were first introduced in \cite{CY02} to provide resistance against adversarial errors in network communications. Since their introduction, NEC codes have been extensively studied; see, for examples, \cite{CY06a,CY06b,YSJL14} and the references in the survey \cite{BMRST13}. 

In this paper, we consider a new type of NEC coding problem introduced and studied in \cite{AW17} and \cite{FKW17}, called \emph{variable packet-error coding} (VPEC). This problem arises in a communication scenario where a source transmits information to a destination over multiple non-intersecting paths in a network (which can be viewed as multiple transmission channels). During transmission, messages sent through these paths may encounter interference, resulting in adversarial errors, which are referred to as \emph{packet errors}. Classical network error-correcting codes typically focus on ensuring robustness against the maximum possible number of packet errors. However, in many real-world systems, such as those where errors arise from adversarial jammers, attacks may occur only sporadically, and the network often operates without errors. This makes the worst-case approach overly pessimistic. Thus, the goal of variable packet-error coding is to design a coding system that meets performance objectives when the maximum number of errors are present while ensuring improved performance when there are fewer or no errors.

In VPEC, one needs to construct codes whose performance dynamically improves as the number of errors decreases. This is not achievable with the conventional approach, where maximum distance separable (MDS) codes with certain minimum distances are typically employed. Since the code parameters are designed to recover the source when the maximum number of errors occur, MDS codes offer no performance improvement when fewer errors are present. Similar problems have also been studied recently for storage codes; see, for example, \cite{KRG19,KMSYRG20,MR22a,MR24}, where the code rate is required to vary over time in response to the variations in failure rates of storage devices. The VPEC problem is also closely related to the multiple-description (MD) problem \cite{Goyal01} in network information theory (see also \cite{GC82,AW12}), and coding approaches in cryptography, such as message authentication codes (MACs) \cite[Sec. 6.1]{Goldreich09} (see also \cite{AW17}). 

%In MD problem, a source is encoded into several messages that are sent to the decoder, only a subset of which reach their destination. Then, the decoder is required to use the subset of encoded messages to reproduce the source, with the fidelity of the reproduction depending on which packets are received. The difference lies in the fact that, in the MD problem, each message is either received correctly or not received at all; the network does not introduce errors.

%MACs assume a computationally-bounded adversary and a shared key that is available to the transmitter and the receiver but not the adversary. VPEC, on the other hand, allows for an omniscient adversary with unbounded computational power.

Based on prior works \cite{AW11} and \cite{AW17}, which showed that source--channel separation is not optimal for the VPEC problem, Fan, Kosut, and Wagner appropriately addressed the VPEC problem using rate--distortion theory in \cite{FKW17}. Specifically, they consider the scenario where a source sequence is encoded into $ N$ packets (or messages) at a given rate $ R $, with up to $ T $ packets potentially being adversarially altered by the network. The decoder receives $ N $ packets without any knowledge of how many or which packets were altered, only knowing that the total number of altered packets is at most $ T $. The decoder then reconstructs the source sequence, and a distortion measure is applied to quantify the discrepancy between the source and its reconstruction. Hence, from this perspective, the goal in VPEC is to design codes that guarantee a specified distortion level when $ T $ errors occur, with progressively lower distortion levels as the number of errors decreases.

As in \cite{FKW17}, we study the VPEC problem under the erasure distortion measure. In this setting, the per-symbol distortion is zero if the reconstruction symbol matches the corresponding source symbol, one if the reconstruction symbol is an ``erasure'' symbol, and infinity otherwise. Hence, as long as the decoder does not incorrectly guess any source symbol, the overall distortion of a sequence is defined as the fraction of erasures in the reconstruction.

%The erasure distortion measure is well-suited for a broad range of sources. For instance, in audio and video applications, interpolation can often compensate for missing samples, pixels, or frames. Similarly, humans can comprehend natural language text even with some erased characters~\cite{ref9}. Even executable computer code, traditionally considered unsuitable for lossy compression, aligns well with the erasure distortion measure: execution could simply pause at erasures, awaiting additional information, without ever executing incorrect instructions. Focusing on the erasure distortion measure also simplifies the problem, serving as a useful starting point for exploring new challenges. This approach mirrors the role of the binary erasure channel in the development of modern coding theory~\cite{ref10}.

In the same spirit as \cite{AW17} and \cite{FKW17}, this paper focuses on the trade-off between the rate $ R $ of the messages encoded in each packet and the (erasure) distortion $ D $ that quantifies the reconstruction accuracy in the VPEC problem. We not only study the general achievability of rate--distortion pairs, but also investigate the performance of rate--distortion pairs for certain classes of codes. Specifically, our contributions are summarized next:
\begin{itemize}
    \item {\bf Bounds on the rate--distortion trade-off:} We establish several lower bounds on the trade-off between the rate $R$ of each packet and the distortion $D$, in terms of the number of packets $N$ and the number of packet errors $T$. These bounds include the first part of Theorem~1 in~\cite{FKW17} as a special case, and one of them is shown to be optimal for certain parameter regimes based on our construction.
    \item {\bf Constructions:} We present two explicit constructions of VPEC codes. The first construction utilizes higher-order MDS codes introduced in \cite{Roth22} (see also \cite{BGM22}) and yields VPEC codes that achieve improved rate--distortion trade-offs compared to the polytope code in \cite{FKW17} for general parameter regimes. The second construction is based on a variant of repetition codes and leads to VPEC codes with $N = 2T + 1$ packets, capable of correcting $T$ packet errors for any positive integer $T$. These codes achieve the bound on the rate--distortion trade-off for $0 \leq D \leq T/N$. Moreover, this construction includes a decoding algorithm with a time complexity of $O(N^3)$.
\end{itemize}

The rest of the paper is structured as follows. In Section~\ref{sec: preliminaries}, we provide a formal description of the VPEC problem along with our main results, and include a comparison with previous works. Section~\ref{sec: bounds} is devoted to proving our bounds on the rate--distortion trade-off for the VPEC problem. In Section~\ref{sec: constructions}, we present our constructions of coding schemes. Finally, Section~\ref{sec: conclusion} concludes the paper by highlighting several open problems for future research.

\section{Problem formulation and main results}\label{sec: preliminaries}

\subsection{Notation}

Let $\Integers$ denote the set of all integers. We use $\Integers^+$ to denote the set of all positive integers, and for $L \in \Integers^{+}$, we use $\Integers / L$ to denote the set of all rational numbers with denominator $L$. For integers $1\le m\le n$, we denote $[m:n]\triangleq \{m,m+1,\ldots,n\}$ and  $[n]\triangleq [1:n]$, and the notation $\Integers_m$ will stand for the ring of integers modulo $m$. For a subset $A$ of $[n]$ and a positive integer $t$, we use ${A\choose \leq t}$ to denote the family of all subsets of $A$ with size at most $t$. Given an alphabet $\Sigma$ of size at least $ 2 $, for words $ \bbx = (x_1, \ldots, x_n) $ and $ \bby = (y_1, \ldots, y_n) $ in $ \Sigma^n $, we use the notation
$$ d_H(\bbx, \bby) \triangleq \abs{\{i \in [n] : x_i \neq y_i\}} $$  
for the Hamming distance between $ \bbx $ and $ \bby $, and we denote by $ B(\bbx, r) $ the Hamming ball in $ \Sigma^n $ centered at $ \bbx $ with radius $ r $. Given a code $\mathcal{C} \subseteq \Sigma^n$, we denote $ d_H(\cC) $ as the minimum distance of the code $ \cC $. For a codeword $ \bbc \in \mathcal{C} $ and a subset $ S \subseteq [n] $, let $ \bbc|_{S} $ denote the word obtained by projecting the coordinates of $ \bbc $ onto $ S $, and define $ \mathcal{C}|_{S} \triangleq \{\bbc|_{S} : \bbc \in \mathcal{C}\} $. Given $d \in [n]$, we call $\cC \subseteq \Sigma^n$ an $(n,d)$-anticode if $d_H(\bbc,\bbc') \le d$ for any $\bbc, \bbc' \in \cC$.

Let $ q $ be a prime power. We use $ \mathbb{F}_q $ to denote the finite field of size $ q $ and $ \mathbb{F}_q^n $ to denote the $ n $-dimensional linear space over $ \mathbb{F}_q $. We call a code $\cC\subseteq \mathbb{F}_q^{n}$ a linear $[n,k]$ code over $\mathbb{F}_q$ if $ \cC $ is a $ k $-dimensional linear subspace of $ \mathbb{F}_q^n $.

% Let $\alpha_1,\alpha_2,\ldots,\alpha_n$, where $n\leq q$, be $n$ pairwise distinct elements of $\mathbb{F}_q$. For $k\in [n]$, denote by $\mathbb{F}_q^{<k}[x]$ the set of polynomials of degree less than $k$. The Reed-Solomon code with evaluation vector $\bm{\alpha}\triangleq (\alpha_1,\alpha_2,\ldots,\alpha_n)\in \Fq^n$ and dimension $k$ is defined by
% \begin{equation}\label{eq_def_RS-code}
%     RS_{\bm{\alpha}}\triangleq \{(f(\alpha_1),f(\alpha_2),\ldots,f(\alpha_n)):f(x)\in \mathbb{F}_q^{<k}[x]\}.
% \end{equation}

Throughout the paper, we use standard asymptotic notation, as follows. Let $f(\cdot)$ and $g(\cdot)$ be two non-negative functions defined on the positive integers. We say that $f=O(g)$ (or $g=\Omega(f)$) if there is some constant $c\geq 0$ such that $f(n)\le cg(n)$ for sufficiently large $n$. %Moreover, unless otherwise specified, all logarithms $\log(\cdot)$ are taken to base-$2$. 

\subsection{Problem formulation and previous results}

Let $\Sigma$ denote the alphabet of the source and let $N$ denote the number of channels, i.e., the number of packets that are sent in each transmission. Given a source sequence $\bbx\in \Sigma^k$ of length $k$, the encoder creates $N$ packets via the following encoding functions
\begin{equation*}
f_{\ell}:~\Sigma^{k}\longrightarrow\Sigma^{kR_{\ell}},~\ell\in [N],
\end{equation*}
for some $R_{\ell} > 0$. We refer to $R_{\ell}$ as the \emph{rate of the $\ell$-th packet ($\ell$-th channel)}.
%, and define the \emph{overall rate} of the encoder (or scheme) as $R = \sum_{\ell \in [N]} R_{\ell}$. 
Then, the encoder sends the following list of $N$ packets
\begin{equation}\label{eq_lists of packets}
    \bbf_N(\bbx) \triangleq (f_1(\bbx),f_{2}(\bbx),\ldots,f_{N}(\bbx)),
\end{equation}
%(This equation will be referred to below by $(C_1,C_2,\ldots,C_N)$.)
and we sometimes use the notation $C_j$ for $f_j(\bbx)$. After transmission through the channels, for any $j \in [N]$, the content $C_j$ of the $j$-th packet may be altered to some different value $\tilde{C}_j \in \Sigma^{kR_{\ell}}$. Let $\tilde{\Sigma}$ denote the reconstruction alphabet, which hereafter will be $\Sigma \cup \{e\}$, where $e$ denotes the erasure symbol. Then, the decoder employs a function
\begin{equation}\label{eq_decoding func}
    g:~\prod_{\ell=1}^{N} \Sigma^{kR_{\ell}}\longrightarrow \tilde{\Sigma}^{k}
\end{equation}
to reproduce the source given the received packets. The fidelity of the reproduction is measured using distortion measure
\begin{align*}
    &\Delta:~\Sigma^{k}\times\tilde{\Sigma}^{k}\rightarrow[0,\infty],\\
    &\Delta(\bbx,\tilde{\bbx})=\frac{1}{k}\sum_{j=1}^{k}\Delta(x_j,\tilde{x}_j),
\end{align*}
where $\Delta(x_j,\tilde{x}_j)$ is the single-letter distortion between $x_j$ and $\tilde{x}_j$. As in \cite{FKW17}, we limit our focus to the \emph{erasure distortion measure} \cite[Page 338]{Information_Theory}: for $x \in \Sigma$ and $\tilde{x} \in \tilde{\Sigma}$,
\begin{equation}\label{eq_erasure_distortion}
    \Delta(x, \tilde{x}) \triangleq
    \begin{cases}
        0, & \text{if } x = \tilde{x}; \\
        1, & \text{if } \tilde{x} = e; \\
        \infty, & \text{otherwise}.
    \end{cases}
\end{equation}

The distortion of the coding scheme defined by the pair $(\bbf_N,g)$ under at most $T$ packet errors is denoted by
\[
D_{T}(\bbf_N,g) \triangleq \max_{\bbx \in \Sigma^k} \max_{A \in \binom{[N]}{\leq T}} \max_{\tilde{C}_A} \Delta\big(\bbx, g(C_{[N]\setminus A}, \tilde{C}_{A})\big).
\]
Here, $g(C_{[N]\setminus A}, \tilde{C}_A)$ denotes the decoder's output when it receives $C_\alpha = f_\alpha(\bbx)$ for all $\alpha \in [N] \setminus A$ and $\tilde{C}_\alpha \in \Sigma^{kR_\alpha}$ for all $\alpha \in A$. We are interested in the rate--distortion trade-off when $T$ can take two different values, $T_0$ and $T_1$.

\begin{definition}[{\cite[Definition 1]{FKW17}}]\label{def_asympto_rate--distortion-vector}
    The rate--distortion vector $(R_1,\ldots,R_N,D_{0},D_{1})$ is \emph{achievable} with respect to $(T_0,T_1)$, if for all $\varepsilon> 0$, there exists a coding scheme $(\bbf_N,g)$ for a source of blocklength $k$, for some $k$, satisfying
    \begin{align}
        \frac{1}{k}\log_{|\Sigma|}|f_{\ell}|&\leq R_{\ell}+\varepsilon,~\ell\in [N]; \nonumber\\
        D_{T_0}(\bbf_N,g)&\leq D_{0}+\varepsilon; \label{eq_def_VPEC_asympto}\\
        D_{T_1}(\bbf_N,g)&\leq D_{1}+\varepsilon.\nonumber
    \end{align}
\end{definition}

In~\cite{FKW17}, Fan, Kosut and Wagner focused on the case where $T_0 = 0$, $T_1 = T$, and all $R_{\ell}$ are equal. Specifically, in their setting, the coding scheme has rate $R$ for each packet, allows lossless reconstruction when there are $T_0 = 0$ packet errors, and ensures an erasure distortion of at most $D$ when there are at most $T_1 = T$ packet errors. We refer to this requirement of the coding scheme as the $T$-VPEC. We say the rate--distortion pair $(R,D)$ is \emph{achievable} for $T$-VPEC if the rate--distortion vector $(R, \ldots, R, 0, D)$ is achievable according to Definition~\ref{def_asympto_rate--distortion-vector} with respect to $(T_0,T_1) = (0,T)$.

Given $N$, $T$, a source of blocklength $k$, and a coding scheme defined by the encoder-decoder pair $(\bbf_N, g)$ of the form~\eqref{eq_lists of packets} and~\eqref{eq_decoding func}, suppose it satisfies~\eqref{eq_def_VPEC_asympto} with parameters $R_{\ell} = R$, $T_0 = D_0 = \varepsilon = 0$, $D_1 = D$, and $T_1 = T$. Let $\mathcal{C}_{(\bbf_N, g)}$ denote the image of $\bbf_N$. We refer to such a set $\mathcal{C}_{(\bbf_N, g)} \subseteq (\Sigma^{kR})^N$ as a \emph{$T$-VPEC code} over $\Sigma$ with parameters $(N, k, R, D)$. When $\Sigma = \mathbb{F}_q$ and each $f_{\ell}$ is a linear map from $\mathbb{F}_q^k$ to $\mathbb{F}_q^{kR}$ for every $\ell \in [N]$, we say that $\mathcal{C}_{(\bbf_N, g)}$ is a linear $T$-VPEC code over $\mathbb{F}_q$. For simplicity, we usually omit the subscript when the coding scheme $(\bbf_N, g)$ is clear from the context. Moreover, in the following sections, we use the notation $\mathcal{C}(\bbx)$ to denote the codeword produced by encoding the message word $\bbx \in \Sigma^k$ using the encoding map $\bbf_N$ that defines $\mathcal{C}$.

The following theorem states the main result from \cite{FKW17}.
\begin{theorem}[{\cite[Theorem 1]{FKW17}}]\label{Fan's_main_result}
    Let $N$ and $T$ be positive integers with $N>T$.
    %, and denote $\theta=T/N$.
    \begin{itemize}
        \item [1)] If $0\leq R<\frac{1}{N-T}$, then there is no finite $D$ for which $(R,D)$ is achievable for $T$-VPEC.
        \item [2)] Let $F(T)$ denote $T+\floorenv{\frac{T^2}{4}}+1$ and suppose that $N\geq F(T)+1$. Then for any $\frac{1}{N-T}\leq R\leq \frac{1}{N-2T}$, the rate--distortion pair
        $$\left(R,\frac{(N-2T)^{-1}-R}{(N-2T)^{-1}-(N-T)^{-1}}\frac{F(T)}{N}\right)$$
        is achievable.
    \end{itemize}
\end{theorem}

In this paper, following the approach of~\cite{FKW17}, we focus on the achievability of rate--distortion pairs $(R,D)$, given $\Sigma$, $N$, and $T$, and assuming a blocklength $k$. 
%Moreover, for given $k$ and $N$, let $\cC$ denote the code in $(\Sigma^{kR})^{N}$ defined by a coding scheme $(f_1, \ldots, f_N, g)$ for a message source of blocklength $k$, satisfying~\eqref{eq_def_VPEC_asympto} with $R_{\ell}=R$, $T_0 = D_0 = \varepsilon = 0$, $T_1 = T$, and $D_1 = D$. We refer to $\cC$ as a \emph{$T$-VPEC code} over $\Sigma$ with parameters $(N, k, R, D)$. When $\Sigma = \mathbb{F}_q$ and $f_{\ell}$ is a linear map from $\mathbb{F}_q^k$ to $\mathbb{F}_q^{kR}$ for every $\ell \in [N]$, we say that $\mathcal{C}$ is a linear $T$-VPEC code over $\mathbb{F}_q$. 
%In this case, $\mathcal{C}$ can be viewed as an array code of size $kR \times N$ over $\mathbb{F}_q$ with dimension $k$. Following \cite{KPLK14} and \cite{SES19}, we can flatten the codewords of $\cC$ into vectors of length $kRN$ by reading the symbols of the codewords column by column, and within each column, from top to bottom. Let $\mathbf{G}$ be the $k \times kRN$ generator matrix for the flattened code. For convenience, in the following sections, we also refer to $\mathbf{G}$ as the generator matrix for $\mathcal{C}$.

\subsection{Higher-order MDS codes}

To present our construction of VPEC codes, we need the notion of higher-order MDS codes defined in~\cite{Roth22}, which builds upon the theory of list decoding.

Let $\code$ be a code in $\Sigma^n$, and let $L, \tau \in \Integers^+$ be given. We say that $\code$ is \emph{$(\tau,L)$-list decodable} if for every $\bby \in \Sigma^n$, there are no $L+1$ distinct codewords of~$\code$ within Hamming distance at most $\tau$ from $\bby$. Given a positive $\tau \in \Integers/(L+1)$, we say that $\code$ is \emph{strongly-$(\tau,L)$-list decodable} (also called \emph{average-radius $(\tau,L)$-list decodable}) if for every $\bby \in \Sigma^n$, the sum of distances from $\bby$ of any $L+1$ distinct codewords of $\code$ exceeds $(L+1)\tau$.  
Note that strong $(\tau,L)$-list decodability implies ordinary $(\lfloor \tau \rfloor,L)$-list decodability.

As a generalization of unique decoding, list decoding was introduced independently by Elias~\cite{Elias57} and Wozencraft~\cite{Wozencraft58} in the 1950s. In contrast to unique decoding, which outputs at most one codeword, list decoding allows multiple candidate codewords. This flexibility enables it to correct more adversarial errors. Over the years, list decoding has found numerous applications in information theory (see \cite{Ahlswede73, Blinovskii86, Blinovsky97, Elias91}) and theoretical computer science (see \cite{CPS99, GUV09, Sivakumar99, STV99, GRS19Essential}).

Let $\code$ be a linear $[n,k]$ code over $\mathbb{F}_q$, let $L \le \min \bigl\{ q-1, \binom{n-1}{k-1} \bigr\}$, and let $\tau \in \Integers/(L+1)$. It was shown in~\cite{Roth22} that $\code$ is strongly-$(\tau,L)$-list decodable only if
\begin{equation}\label{eq:stronglySingletonImproved}
\tau \le \frac{L(n-k)}{L+1}.
\end{equation}
A linear code is said to be \emph{$L$-MDS} if it attains the bound in~\eqref{eq:stronglySingletonImproved}, and \emph{$\Int{L}$-MDS} if it is $L'$-MDS for every $L' \in \Int{L}$. As shown in~\cite{Roth22}, every $L$-MDS code is also an MDS code.

\subsection{Main results}

Our first result is the following theorem, which provides several bounds on the trade-off between the rate $R$ of each packet and the distortion $D$ for the $T$-VPEC problem.

\begin{theorem}\label{main_Thm1}
    Let $N$ and $T$ be positive integers with $N > T$. Then, the following holds.
    \begin{itemize}
        \item [1.] If the rate--distortion pair $(R,D)$ is achievable for $T$-VPEC, then we have 
        \begin{equation}\label{eq_bound1}
        R\geq 
        \begin{cases}
            \frac{1}{N-T},&~\text{$0\leq D\leq 1$};\\
            \frac{1}{N-2T}, &~\text{$D= 0$}.
        \end{cases}
        \end{equation}
        Moreover, $D=0$ holds only if $N\geq 2T+1$. 
        \item [2.] Suppose that $N \geq 2T + 1$ and let $Ant_q(k,d)$ denote the maximum size of a $(k,d)$-anticode over $\Sigma$ with $|\Sigma| = q$. Given $0\leq D<1$, if there is a $T$-VPEC code over $\Sigma$ with parameters $(N,k,R,D)$, then we have 
        \begin{equation}\label{eq_bound2}
            R\geq \left(1-\frac{\log_{q}(Ant_q(k,kD))}{k}\right)\frac{1}{N-2T}.
        \end{equation}
        \item [3.] Suppose that $N\geq 2T+1$ and that there exists a linear $T$-VPEC code $\mathcal{C}$ over $\mathbb{F}_{q}$ with parameters $(N,k,R,D)$, where $D \in \Integers/k$. Then, we have 
        $$
        R\geq \frac{1-D}{N-2T}.
        $$
    \end{itemize}
\end{theorem}

Note that the bound on $R$ in~\eqref{eq_bound1} of Theorem~\ref{main_Thm1} contains the result from the first part of Theorem~\ref{Fan's_main_result}. Moreover, the value of $Ant_q(k,d)$ is characterized by the well-known Diametric Theorem of Ahlswede and Khachatrian~\cite{AK98}. Therefore, when $q \geq 2(k - kD)/3 + 2$, a closed-form expression for the bound in the second part of Theorem~\ref{main_Thm1} can be derived. This is stated in the following corollary; see Appendix~\ref{sec: Diametric Thm and proof of Coro1} for a detailed description of the Diametric Theorem and the proof of the corollary.

\begin{corollary}\label{coro1}
    Given $0\leq D\leq 1$, let $k$, $N$, $T$, and $q$ be positive integers such that $N\geq 2T+1$ and $q\geq 2k(1-D)/3+2$. If there is a $T$-VPEC code over $[q]$ with parameters $(N,k,R,D)$, then 
        \begin{equation}\label{eq3_coro1}
            R\geq \max\left\{\frac{1-D}{N-2T},\frac{1}{N-T}\right\}.
        \end{equation}
\end{corollary}

%The third part of Theorem \ref{main_Thm1} and the rate--distortion of random-linear codes.

Our second result includes two explicit constructions of VPEC codes. The first construction employs $L$-MDS codes and yields $T$-VPEC codes that achieve improved rate--distortion trade-offs compared to existing results for general parameters $N \geq 2T + 1$. The construction is detailed in Construction~\ref{Cons_L-MDS} in Section~\ref{sec: construction by L-MDS}, and the rate--distortion performance of the resulting $T$-VPEC code is summarized in the following theorem.

\begin{theorem}\label{Thm_Cons_L-MDS}
Let $N$, $T$ and $L$ be positive integers such that $N \geq 2T + 1$, $2 \leq L \leq N/T$ and $L\mid T$. If there exists an $[N, \rho N]$ $L$-MDS code with
\begin{equation}\label{eq_delta}
    \rho = 1 - \left( 1 + \frac{1}{L} \right) \frac{T}{N},
\end{equation}
then there is an explicit construction of $T$-VPEC code with parameters
\[
\left(N,\rho N^2,\frac{1}{\rho N},\frac{LT}{N}\right).
\]
\end{theorem}

Using a similar time-sharing argument as in~\cite{FKW17}, Theorem~\ref{Thm_Cons_L-MDS} yields the following achievable rate--distortion pairs.

\begin{corollary}\label{coro_Cons_L-MDS}
Let $N$, $T$ and $L$ be positive integers such that $N \geq F(T) + 1$, $2 \leq L \leq N/T$ and $L\mid T$, where $F(T) = T + \left\lfloor \frac{T^2}{4} \right\rfloor + 1$. If there exists an $[N, \rho N]$ $L$-MDS code with $\rho$ defined as in \eqref{eq_delta}, then for any $\frac{1}{N-T} \leq R \leq \frac{1}{N-2T}$, the rate--distortion pair $(R, D)$ is achievable for any $D \leq 1$ satisfying
\[
D \geq \frac{1-T/N - \rho(N-T) R}{N-T - \rho N} \cdot F(T) + \frac{\rho(N-T) R - \rho}{N-T - \rho N} \cdot L T,
\]
when $R \leq (\rho N)^{-1}$, and
\[
D \geq \frac{\rho - \rho(N - 2T) R}{\rho N-(N - 2T)} \cdot L T,
\]
when $R > (\rho N)^{-1}$.
\end{corollary}

\begin{remark}\label{rmk_comparsion_2-MDS}
    \begin{enumerate}
        \item When $R = (N - (1+1/L)T)^{-1}$, the distortion achieved by Theorem \ref{Fan's_main_result} is given by
        \begin{align*}
            & \frac{(N-2T)^{-1}-R}{(N-2T)^{-1}-(N-T)^{-1}} \cdot \frac{F(T)}{N} \\
            = & \frac{(L-1)(N-T)}{LN-(L+1)T} \cdot \frac{F(T)}{N} \\
            = & \frac{(L-1)(N-T)}{L(N-T)-T} \cdot \frac{F(T)}{LT} \cdot \frac{LT}{N}.
        \end{align*}
        This is strictly larger than $LT/N$ when $T \geq 4(L+1)$, as
        \begin{align*}
            \frac{(L-1)(N-T)}{L(N-T)-T} \cdot \frac{F(T)}{LT} & > \frac{L-1}{L}\cdot \frac{F(T)}{LT} \\
            & \geq \frac{L-1}{L}\cdot \frac{1+\frac{T}{4}}{L} \\
            & \geq \frac{(L-1)(L+2)}{L^2}>1,
        \end{align*}
        where the second inequality follows by $F(T) \geq T + \frac{T^2}{4}$, and the third inequality follows by $T \geq 4(L+1)$. Thus, Theorem \ref{Thm_Cons_L-MDS} provides a better achievable rate--distortion pair than that in Theorem \ref{Fan's_main_result} for $T \geq 4(L+1)$.        
        \item In \cite{Roth22}, explicit constructions of $ \left[N,N-3T/2\right] $ $ 2 $-MDS codes over fields of size $ q = O(N^{(3T/2)^{3T}}) $ were provided, which is polynomial in $ N $ for a fixed $ T $. Whereas the polytope code construction in \cite{FKW17} relies on infinite fields, specifically the reals. Thus, Theorem \ref{Thm_Cons_L-MDS} imposes a more relaxed requirement on the field size compared to that in \cite{FKW17}. 
        % \item As detailed later in Construction \ref{Cons_L-MDS} in Section \ref{sec: constructions}, our construction allows the code length to grow linearly with the number of packet errors $ T $. Whereas the polytope code construction in \cite{FKW17} requires $ n \geq F(T) + 1 $, where $ F(T) $ grows quadratically with $ T $.
    \end{enumerate}
\end{remark}

\begin{figure}[htbp]
    \centering    
    \includegraphics[width=0.5\textwidth]{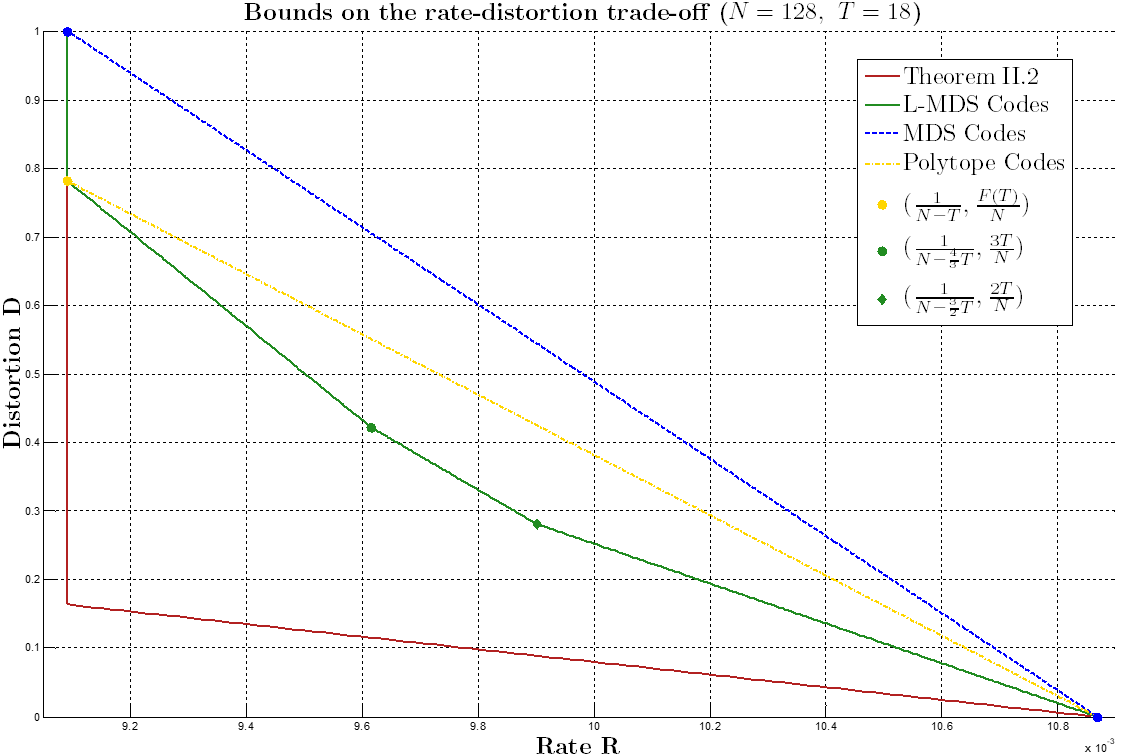}
    \caption{Comparison of the results on the rate--distortion trade-off from~\cite{FKW17}, Theorem~\ref{main_Thm1}, and Theorem~\ref{Thm_Cons_L-MDS}, for $N = 128$ and $T = 18$.}
    \label{figure2}
\end{figure}

In Figure \ref{figure2}, we compare our bounds on the rate--distortion trade-off from Theorem \ref{main_Thm1} and the result of Theorem \ref{Thm_Cons_L-MDS} with known results in \cite{FKW17} for the case $ N=128,~T=18 $, and it has the following features:
%when $ q \geq n^{\left(\frac{3T}{2}\right)^{3T}} $. 

\begin{itemize}
    \item The blue line is obtained by connecting the points $\left((N-2T)^{-1}, 0\right)$ and $\left((N-T)^{-1}, 1\right)$, which correspond to the rate--distortion pairs achieved by $[N, N-2T]$ and $[N, N-T]$ MDS codes, respectively. As shown in~\cite{FKW17}, the rate--distortion pairs corresponding to other points on this line is achieved via time-sharing. Specifically, by dividing the message to be transmitted into $ M $ sub-messages, each containing an equal number of bits, one can, for any $ 0 < c < 1 $, use an $[N, N - 2T]$ MDS code to encode the first $ cM $ sub-messages and an $[N, N - T]$ MDS code to encode the remaining $ (1 - c)M $ sub-messages. This results in a coding scheme that achieves the rate--distortion pair $ \left(\frac{c}{N - 2T} + \frac{1 - c}{N - T},1 - c  \right) $.
    
    \item The yellow line corresponds to the second part of Theorem~\ref{Fan's_main_result}, and is obtained by connecting the points $ \left( (N - T)^{-1}, F(T)/N \right) $ and $ \left( (N - 2T)^{-1}, 0 \right) $. In fact, Fan, Kosut and Wagner \cite{FKW17} showed that the rate--distortion pair $ \left( (N - T)^{-1}, F(T)/N \right) $ is achieved by polytope codes for $ N \geq F(T)+1 $. The rate--distortion pairs corresponding to other points are also achieved through time-sharing arguments.
    
    \item The dark red line corresponds to our bound from Theorem~\ref{main_Thm1} for linear $T$-VPEC codes. The dark green line corresponds to the result of Theorem~\ref{Thm_Cons_L-MDS}, obtained by connecting the following points:
    \[
    \left(\frac{1}{N - T}, \frac{F(T)}{N}\right),\quad \left(\frac{1}{N - \frac{4}{3}T}, \frac{3T}{N}\right),\quad \left(\frac{1}{N - \frac{3}{2}T}, \frac{2T}{N}\right),\quad \text{and} \quad \left(\frac{1}{N - 2T}, 0\right).
    \]
    Among these, the first and last points correspond to the rate--distortion pairs achieved by the polytope codes in~\cite{FKW17} and the $[N, N - 2T]$ MDS codes, respectively. The parameters $T = 18$ and $N = 128$ satisfy
    \[
    N \geq F(T) + 1 = 101,\quad N/T > 3,\quad \text{and} \quad 2, 3 \mid T,
    \]
    which meet the conditions of Theorem~\ref{Thm_Cons_L-MDS} for both $L = 2$ and $L = 3$. Therefore, the second and third points represent the rate--distortion pairs achieved by the $T$-VPEC codes in Theorem~\ref{Thm_Cons_L-MDS} (see Construction~\ref{Cons_L-MDS}) obtained using $2$-MDS and $3$-MDS codes, respectively. Clearly, the dark green line achieves a better rate--distortion trade-off than that achieved by the polytope codes in~\cite{FKW17}.
\end{itemize}

\begin{remark}\label{rmk_asympto}
For a coding scheme defined by an encoder-decoder pair $(\mathbf{f}_{N}, g)$ of the form~\eqref{eq_lists of packets} and~\eqref{eq_decoding func}, we define its \emph{overall rate} as the sum of the per-packet rates:
$$
R^{O} \triangleq \sum_{i=1}^{N} R_{i}.
$$
In the case of $T$-VPEC, where each packet is required to have the same rate, we have $R^{O} = N R$. We say that a pair $(R^{O}, D)$ is \emph{achievable} for $T$-VPEC if the rate-distortion pair $(R, D)$ is achievable, and the converse holds naturally.

As in classical coding problems, we can study the asymptotic behavior of achievable pairs $(R^O, D)$ in the regime where the error fraction $T/N = \theta$ is fixed for some constant $0 < \theta < 1$, and $N \to \infty$. In this setting, the result for polytope codes from Theorem~\ref{Fan's_main_result} becomes trivial, whereas $L$-MDS codes continue to outperform MDS codes. For detailed comparisons, we refer interested readers to Appendix~\ref{sec: asympto}.
\end{remark}

Our second construction employs a variant of the repetition code and yields a $T$-VPEC code with an optimal rate--distortion trade-off in the case where $N = 2T+1$ and $(2T+1) \mid k$. The code construction is presented in Construction~\ref{Cons2} in Section~\ref{sec: construction with opt trade-off}, and the rate--distortion performance of the resulting $T$-VPEC code is summarized in the following theorem.

\begin{theorem}\label{thm_cons2_s=T}
Let $ k $, $ s $ and $ T $ be positive integers satisfying $ (2T+1) \mid k $ and $s\leq T$. Then, for any given alphabet $\Sigma$, there is a $ T $-VPEC code over $\Sigma$ with parameters 
$$ \left(2T+1, k, 1 - \frac{s}{2T+1}, \frac{s}{2T+1}\right) $$ 
and a decoding algorithm with a time complexity of $ O(T^3) $.
\end{theorem}

As an immediate corollary, Theorem \ref{thm_cons2_s=T} provides the following achievable rate--distortion pair.

\begin{corollary}\label{coro_cons2_s=T}
Given positive integers $T$, for any $(T+1)/(2T+1) \leq R \leq 1$, the rate--distortion pair $(R, 1 - R)$ is achievable for the $T$-VPEC problem with $2T + 1$ packets.
\end{corollary}

\begin{remark}\label{rmk_comparsion_Cons1}
Note that for any $ 0 \leq D \leq s/(2T+1) $, it always holds that  
\begin{align*}
    1 - D &\geq 1 - \frac{s}{2T+1} \\
          &> \frac{1}{2} \geq \frac{1}{T+1}.
\end{align*}
For $0 \le D \le T/(2T+1)$ and $k = N = 2T+1$, the bound in Corollary \ref{coro1} implies that
\[
R \ge 1 - D ,
\]
whenever $|\Sigma| \ge 2k/3 + 2 = 4(T+2)/3$. This confirms that the $ T $-VPEC code in Theorem \ref{thm_cons2_s=T} achieves the optimal rate--distortion trade-off when $|\Sigma| \ge 4(T+2)/3$.
\end{remark}

\begin{figure}[htbp]
    \centering
    \subfigure{
    \includegraphics[width=0.44\textwidth]{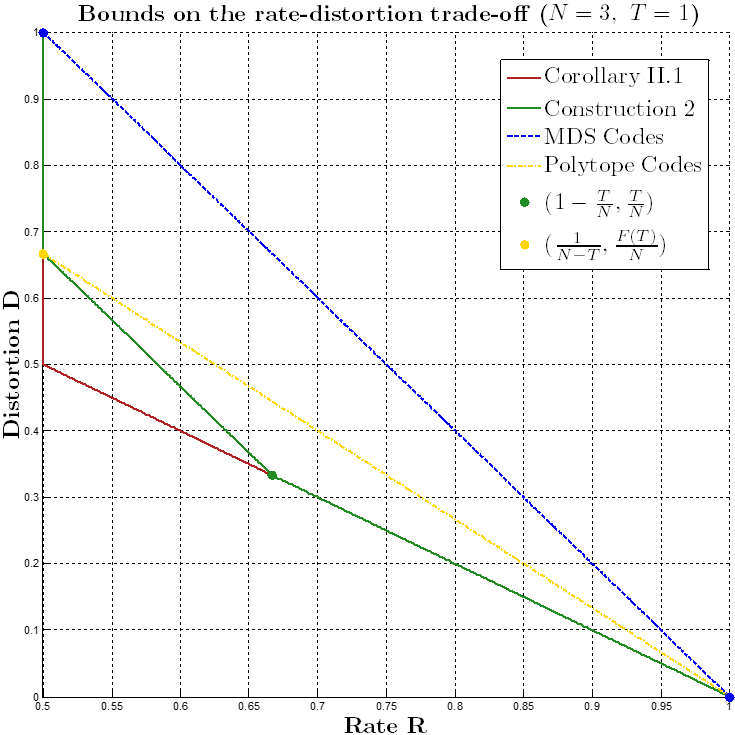}
    }    
    \subfigure{
    \includegraphics[width=0.44\textwidth]{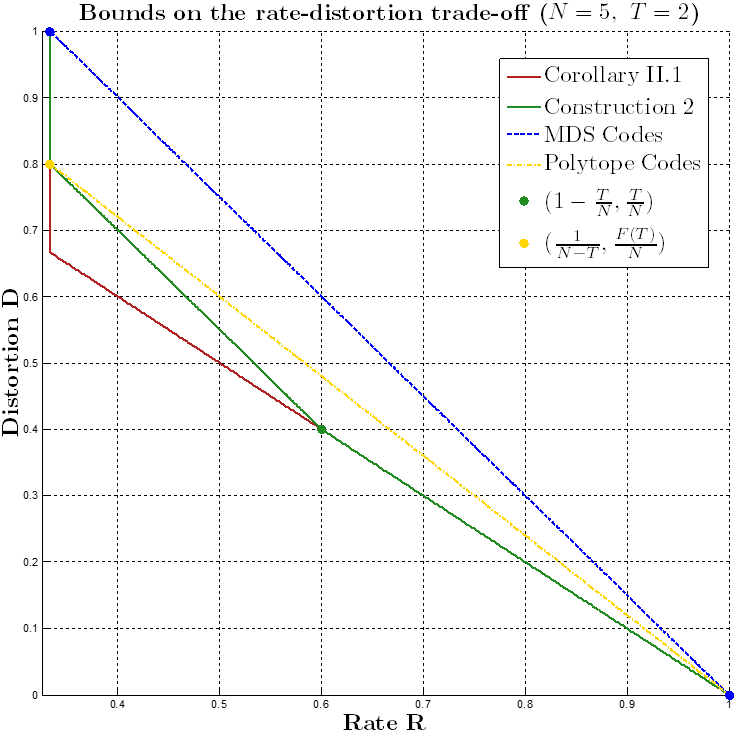}
    }
    \caption{Comparison of the results on the rate--distortion trade-off from~\cite{FKW17}, Corollary~\ref{coro1}, and Theorem~\ref{thm_cons2_s=T}, for $N = 3$, $T = 1$ and $N = 5$, $T = 2$.}
    \label{figure1}
\end{figure}

In Figure \ref{figure1}, we compare our bound in Corollary \ref{coro1} and the result of Theorem \ref{thm_cons2_s=T} with known results in \cite{FKW17} for the cases $ N=3,~T=1 $ and $ N=5,~T=2 $. The sub-figures in Figure \ref{figure1} have the following common features:

\begin{itemize}
    \item As in Figure~\ref{figure2}, the dark red line corresponds to the rate--distortion bound from Corollary~\ref{coro1}, the blue and yellow lines correspond to the rate--distortion pairs achieved by MDS codes and the polytope codes in~\cite{FKW17}, respectively.
    \item The dark green line corresponds to the result of Theorem~\ref{thm_cons2_s=T}. In the regime where $(N-T)/N \leq R \leq 1$, it coincides with the dark red line. Since Theorem~\ref{thm_cons2_s=T} guarantees achievability only for distortions up to $D \leq T/(2T+1)$, the segment of the dark green line corresponding to larger distortions (i.e., $D > T/(2T+1)$, or equivalently $R < (T+1)/(2T+1)$) is obtained by connecting the points $\left((T+1)^{-1}, F(T)/(2T+1)\right)$ and $\left((T+1)^{-1}, T/(2T+1)\right)$. These two points represent the rate--distortion pairs achieved by the polytope codes in~\cite{FKW17} and the $T$-VPEC code in Theorem~\ref{thm_cons2_s=T} (see Construction~\ref{Cons2}), respectively.
\end{itemize}

\section{Bounds on the rate--distortion trade-off}\label{sec: bounds}

In this section, we present the proof of Theorem \ref{main_Thm1}. Let $q$, $k$, $T$, and $n > T$ be positive integers, and let $|\Sigma| = q$. For $0 \leq R \leq 1$, denote $Q \triangleq \Sigma^{kR}$. We start with the following lemma which characterizes the necessary and sufficient conditions for $ \cC $ being a $ T $-VPEC code.

\begin{lemma}\label{lem_VPEC-code_property1}
Let $\mathcal{C}$ be the image of an encoder $\bbf_N: \Sigma^{k} \longrightarrow Q^{N}$. Then, $\mathcal{C}$ is a $T$-VPEC code with parameters $(N, k, R, D)$ for $0 \leq D \leq 1$ if and only if the following conditions hold:
\begin{itemize}
    \item[1.] The minimum Hamming distance of $\mathcal{C}$ satisfies $d_H(\mathcal{C}) \geq T + 1$.
    \item[2.] 
    For any $\bby \in Q^N$, there exists a subset $S \subseteq [k]$ of at least $(1 - D)k$ coordinates with the property that for any two codewords $\mathcal{C}(\bbx)$ and $\mathcal{C}(\tilde{\bbx})$ in $B(\bby, T)$, corresponding to message words $\bbx, \tilde{\bbx} \in \Sigma^k$, it holds that $\bbx|_S = \tilde{\bbx}|_S$.
\end{itemize}
\end{lemma}

\begin{IEEEproof}
    We begin with the proof of the ``if'' part. Given a code $ \mathcal{C} \subseteq Q^N $ satisfying conditions 1 and 2, we define the decoder $ g: Q^N \to \tilde{\Sigma}^k $ as follows:  for any received word $ \bby \in Q^N $, suppose that $ \mathcal{C} \cap B(\bby, T) = \{\mathcal{C}(\bbx_1), \mathcal{C}(\bbx_2), \ldots, \mathcal{C}(\bbx_m)\} $ for some $ \bbx_1, \bbx_2, \ldots, \bbx_m \in \Sigma^k $, then  
    \begin{equation*}
        g(\bby)(j) =
    \begin{cases} 
    \bbx_1(j), & \text{if } j \in S; \\ 
    e, & \text{otherwise},
    \end{cases}
    \end{equation*}
    where $S\triangleq \{j\in [k]: \bbx_1(j) = \bbx_2(j) = \cdots = \bbx_m(j)\}$ is the set of coordinates that all $\bbx_i$s agree.

    If $\bby \in \mathcal{C}$, i.e., $\bby = \cC(\bbx_0)$ for some $\bbx_0 \in \Sigma^k$. Since $d_{H}(\cC) \geq T + 1$, it follows that $\cC \cap B(\bby, T) = \{\cC(\bbx_0)\}$. Thus, $g(\bby) = \bbx_0$, and we have $\Delta(\bbx_0, g(\bby)) = 0$ in this case.
    
    If $ \bby \notin \cC $, then $ \bby $ can only be obtained from $ \cC(\bbx_1), \ldots, \cC(\bbx_m) $ by altering at most $ T $ packets. Then, by condition 2, $ \abs{S}\geq (1 - D)k $. Thus, for each $ i \in [m] $, we have $ \Delta\left(\bbx_i, g(\bby)\right) \leq D $. Hence, the encoder-decoder pair $(\bbf_{N},g)$ satisfies \eqref{eq_def_VPEC_asympto} with parameters $R_{\ell} = R$, $T_0 = D_0 = \varepsilon = 0$, $D_1 = D$, and $T_1 = T$.
    By definition, $ \mathcal{C} $ is a $ T $-VPEC code with parameters $ (N, k, R, D) $. This confirms the ``if'' part.

    Next, suppose that $ \mathcal{C} $ is a $ T $-VPEC code with parameters $ (N, k, R, D) $ defined by an encoder-decoder pair $(\bbf_{N},g)$, and we proceed with the proof for the ``only if'' part.

    Assume, towards a contradiction, that $d_{H}(\cC)\leq T$. Then, there exist two distinct messages words $\bbx, \tilde{\bbx} $ such that the Hamming distance between their encodings satisfies $ d_H(\mathcal{C}(\bbx), \mathcal{C}(\tilde{\bbx})) \leq T $. Suppose that the codeword $ \mathcal{C}(\bbx) $ was transmitted, but the codeword $ \mathcal{C}(\tilde{\bbx}) $ was received. Since $ \mathcal{C} $ allows lossless reconstruction when no packet errors occur, it follows that $ g(\mathcal{C}(\tilde{\bbx})) = \tilde{\bbx} $. However, by \eqref{eq_erasure_distortion}, this implies  
    \[
    \Delta(\bbx, g(\mathcal{C}(\bbx))) = \Delta(\bbx, \tilde{\bbx}) = \infty,
    \]  
    which contradicts the assumption that $ D \leq 1 $. This shows that $\cC$ satisfies condition 1.

    Next, fix $ \mathbf{y} \in Q^N $. By definition, $ g(\mathbf{y}) \in \tilde{\Sigma}^{k} $. For any $\mathcal{C}(\bbx) \in B(\mathbf{y}, T)$, since $\mathcal{C}$ is a $T$-VPEC code, it holds that $ \Delta(\bbx, g(\mathbf{y})) \leq D $. In particular, $\bbx$ agrees with $g(\mathbf{y})$ on all those coordinates where $g(\mathbf{y})$ is not equal to the erasure symbol $e$, and the number of such coordinates is at least $(1 - D)k$. This shows that $\cC$ satisfies condition 2 and the ``only if'' part follows.
\end{IEEEproof}

\begin{corollary}\label{Coro_VPEC-code_distance}
    Suppose that $\cC$ is a $T$-VPEC code with parameters $(N,k,R,D)$. Then 
    \[
    d_{H}(\cC)\geq
    \begin{cases}
        T+1,&~0<D\leq 1;\\
        2T+1,&~D=0.
    \end{cases}
    \]
\end{corollary}

\begin{IEEEproof}
It remains to prove the case when $ D = 0 $.

Suppose $ D = 0 $. By the condition 2 in Lemma \ref{lem_VPEC-code_property1}, the ball $ B(\mathbf{y}, T) $, for any $ \mathbf{y} \in Q^N $, contains at most one codeword. This immediately implies that the balls of radius $ T $ centered at the codewords are disjoint. Consequently, the minimum distance of the code must be at least $ 2T + 1 $.
\end{IEEEproof}

\begin{IEEEproof}[Proof of Theorem \ref{main_Thm1}]
    Let $\cC$ be a $T$-VPEC code in $Q^N$ with parameters $(N,k,R,D)$ defined by an encoder-decoder pair $(\bbf_{N},g)$.

    By the Singleton bound, we have 
    \begin{align*}
        q^k=|\cC|\leq |Q|^{N-d_{H}(\cC)+1}=q^{kR(N-d_{H}(\cC)+1)}.
    \end{align*}
    This implies that $R\geq (N-d_{H}(\cC)+1)^{-1}$. Then, part $1$ of the theorem follows directly by Corollary \ref{Coro_VPEC-code_distance}.
    
    Next, we proceed to the proof of the second result. 
    
    Since $N\geq 2T+1$, then, by the pigeonhole principle, $\cC$ has a subset of codewords $\cC'$ of size at least $$\frac{\abs{\cC}}{|Q|^{N-2T}}$$ 
    such that they all agree on their first $N-2T$ coordinates. Let $\cC(\bbx),\cC(\tilde{\bbx})\in \cC'$ be two such codewords. Then there exists a word $\bby\in Q^{N}$ such that $d_{H}(\bby,\cC(\bbx)),d_{H}(\bby,\cC(\tilde{\bbx}))\leq T$. By Lemma \ref{lem_VPEC-code_property1}, $\bbx$ and $\tilde{\bbx}$ agree on at least $(1-D)k$ coordinates. Thus, $d_{H}(\bbx, \tilde{\bbx})\leq kD$. This implies that the set of message words 
    $$\{\bbx\in \Sigma^{k}: \cC(\bbx)\in\cC'\}$$
    is a $(k,kD)$-anticode. Thus, we have
    $|\cC'|\leq Ant_q(k,kD)$. Then, part $2$ follows directly by 
    $\abs{\cC'}\geq\frac{\abs{\cC}}{|Q|^{N-2T}}$ and $|\cC|=q^k$.

    Finally, we proceed to the proof of part $3$.

    Recall that the subcode $\cC'$ consists of all the codewords in $\cC$ that agree on the first $N-2T$ coordinates. Let $\bbc_0\in \cC'$. Then, denote
    \[
    \cC''\triangleq\{\bbc-\bbc_0:~\bbc\in \cC'\}.
    \]
    Since $\cC$ is linear, $\cC''$ is the subcode of $\cC$ consisting of all codewords in $\cC$ that are zero on the first $N - 2T$ coordinates. Clearly, $\cC''$ is a linear subcode of $\cC$ in $(\mathbb{F}_{q}^{kR})^{N}$. This implies that the set $\{\bbx \in \mathbb{F}_q^{k} : \cC(\bbx) \in \cC''\}$ is a linear subspace of $\mathbb{F}_q^{k}$.

    On the other hand, according to the proof of part $2$, the set $\{\bbx \in \mathbb{F}_q^{k} : \cC(\bbx) \in \cC''\}$ is also a $(k, kD)$-anticode. Hence, its dimension is at most $kD$. Noting that $|\cC''| = |\cC'|$, we have
    \begin{align*}
    q^{kD} &\geq \abs{\{\bbx \in \mathbb{F}_q^{k} : \cC(\bbx) \in \cC'\}} \\
           &= \abs{\cC'} \geq q^{k - (N - 2T)kR},
    \end{align*}
    which leads to the inequality $R \geq \left(1 - D\right)/(N - 2T)$.

    This completes the proof.
\end{IEEEproof}

\section{Constructions of VPEC coding schemes}\label{sec: constructions}

In this section, we present two explicit constructions of $ T $-VPEC codes. The first construction employs $L$-MDS codes together with an interleaving technique, resulting in a $T$-VPEC code that achieves a better rate--distortion pair than the polytope code in~\cite{FKW17}. The second construction yields $T$-VPEC codes of length $N = 2T + 1$ that achieve the optimal rate--distortion trade-off, along with a decoding algorithm of time complexity $O(N^3)$.

%%%%%%%%%%%%%%%%%%%%%%%%%%%%%%%%%%%%%%%%%%%%%%

\subsection{VPEC codes from Higher-Order MDS codes}\label{sec: construction by L-MDS}

In this section, we prove Theorem~\ref{Thm_Cons_L-MDS} by constructing $T$-VPEC codes using $L$-MDS codes. To start, we introduce the notion of \emph{interleaved codes}.

Given $\ell, n \in \Integers^+$, for $i \in \Int{\ell}$ let $\code_i$ be a code in $\Sigma_i^n$. The \emph{interleaved code} $\otimes_{i \in \Int{\ell}} \code_i$ over the alphabet $\otimes_{i \in \Int{\ell}} \Sigma_i = \Sigma_1 \times \Sigma_2 \times \cdots \times \Sigma_\ell$ is defined as the set
\[
\Bigl\{ \Gamma = (c_{i,j})_{i \in \Int{\ell}, j \in \Int{n}} \,:\,
(c_{i,j})_{j \in \Int{n}} \in \code_i,
\; \textrm{for $i \in \Int{\ell}$}
\Bigr\} ,
\] 
namely, it consists of all $\ell \times n$ arrays~$\Gamma$ in which each row~$i$ is a codeword in~$\code_i$ and each column is viewed as a symbol in $\otimes_{i \in \Int{\ell}} \Sigma_i$. When all the constituent codes $\code_i$ are the same code~$\code$, we use the notation $\code^{\otimes \ell}$ and call it the \emph{$\ell$-level interleaving} of~$\code$. If $\code$ is a linear $[n,k]$ code over $\mathbb{F}_q$ with a generator matrix~$\bbG$ and parity-check matrix~$\bbH$, then we can view $\code^{\otimes \ell}$ as a linear $[n,k]$ code over the extension field $\Phi \triangleq \mathbb{F}_{q^{\ell}}$:
We do this by interpreting each column in an $\ell \times n$ code array~$\Gamma$ as the representation of an element of~$\Phi$, expressed as a column vector in $\mathbb{F}_q^\ell$ with respect to some fixed basis of~$\Phi$ over~$\mathbb{F}_q$. The matrices~$\bbG$ and~$\bbH$ then function as the generator matrix and parity-check matrix, respectively, of the code $\code^{\otimes \ell}$.

\begin{proposition}\label{prop:lifting}
Let $\code$ be a linear $[n,k]$ code over $\mathbb{F}_q$ and suppose that $\code$ is $(\tau,L)$-list decodable, for some $L, \tau \in \Integers^+$. If $q > \binom{L+1}{2}$ then $\code^{\otimes \ell}$ is $(\tau,L)$-list decodable, for every $\ell \in \Integers^+$.
\end{proposition}

\begin{IEEEproof}
Suppose, to the contrary, that $\code^{\otimes \ell}$ is not $(\tau, L)$-list decodable. That is, there exists an array $\bbY \in \mathbb{F}_q^{\ell \times n}$ and $L + 1$ distinct codewords $\Gamma_1, \Gamma_2, \ldots, \Gamma_{L+1} \in \code^{\otimes \ell}$ such that their Hamming distances satisfy
\[
d_H(\bbY, \Gamma_r) \le \tau, \quad \text{for all } r \in \Int{L+1},
\]
when viewed as vectors of length $n$ over $\mathbb{F}_{q^{\ell}}$. Then for every row vector $\bbu \in \mathbb{F}_q^{\ell}$,
\[
d_{H}(\bbu \cdot \bbY, \bbu \cdot \Gamma_r) \le \tau ,
\quad \textrm{for $r \in \Int{L+1}$} .
\]
Since $\bbu \cdot \bbY \in \mathbb{F}_q^n$ and $\bbu \cdot \Gamma_r \in \code$ for all $r \in \Int{L+1}$, we will reach a contradiction once we find some $\bbu \in \mathbb{F}_q^\ell$ for which the codewords $\bbu \cdot \Gamma_r$ are distinct for all $r \in \Int{L+1}$. Indeed, for any two distinct $r, s \in \Int{L+1}$ we have $\bbu \cdot \Gamma_r = \bbu \cdot \Gamma_s$, if and only if $\bbu$ belongs to the left kernel of (the nonzero matrix) $\Gamma_r - \Gamma_s$; the size of this kernel, in turn, is at most $q^{\ell-1}$. Ranging over all unordered pairs $\{ r, s \} \subseteq \Int{L+1}$, by a simple union bound argument, there exist at least 
\[
q^\ell - \binom{L+1}{2} q^{\ell-1} > 0
\]
vectors $\bbu \in \mathbb{F}_{q^{\ell}}$ that belong to none of these kernels. Therefore, the result follows since $q > \binom{L+1}{2}$.
\end{IEEEproof}

A $(\tau, L)$-list decoder for the code $\code^{\otimes \ell}$ in Proposition~\ref{prop:lifting} (which decodes any pattern of up to $\tau$ column errors by returning a list of codewords of size ${}\le L$) can be obtained by iteratively applying a $(\tau, L)$-list decoder for~$\code$ on the rows of a received array $\bbY \in \mathbb{F}_q^{\ell \times n}$, provided that $\bbY$ contains at most $\tau$ column errors. Indeed, assume by induction on~$\ell$ that we are able to decode the $(\ell-1) \times n$ array $\bbY'$ formed by the first $\ell-1$ rows of~$\bbY$, ending up with at most $L$~codewords $\Gamma'_1, \Gamma'_2, \ldots \in \code^{\otimes (\ell-1)}$. We then decode the $\ell$th row of~$\bbY$ using a $(\tau,L)$-list decoder for~$\code$, ending up with at most $L$ codewords $\bbc_1, \bbc_2, \ldots \in \code$. Among all the (at most) $L^2$ codewords of $\code^{\otimes \ell}$ of the form
\[
\Gamma_{r,s}
=
\left(
\begin{array}{c}
\Gamma'_r \\ \hline \bbc_s
\end{array}
\right)
\in \code^{\otimes \ell},
\quad r, s \in \Int{L},
\]
we are guaranteed that there are at most $L$ pairs $(r, s)$ for which $d_H(\bbY, \Gamma_{r,s}) \le \tau$.

\begin{remark}
Regarding the list decoding of generic interleaved codes, the only work we are aware of is \cite{GGR2011}. In \cite{GGR2011}, it was shown that the list decoding radius of $\mathcal{C}^{\otimes \ell}$ remains unchanged compared to that of $\mathcal{C}$; see Theorem 2.5 in \cite{GGR2011} for a detailed description. This result does not imply Proposition~\ref{prop:lifting}. In fact, the proof of Lemma 3.7 in \cite{GGR2011} shows that Proposition~IV.1 is generally false if we do not put any restrictions on the field size $q$ (in terms of $L$). For example, if we take $\code$ to be the $[n,1]$ repetition code over $\mathbb{F}_q$, then $\code$ is clearly $(n, L)$-list decodable for $L = q$, but its $\ell$-level interleaving is not $(n, L')$-list decodable for any $L' < q^\ell$.
\end{remark}

\begin{remark}
Following a similar proof, one can show that the result of Proposition~\ref{prop:lifting} also holds when $\mathcal{C}$ is an $\mathbb{F}_q$-linear code over a finite vector space $\Sigma$ defined over $\mathbb{F}_q$, provided that $q > \binom{L+1}{2}$. Moreover, we show that strong $(\tau, L=2)$-list decodability is preserved under interleaving, where $\tau$ is of the form $a/(L+1)$ for some $a \in \mathbb{Z}^+$. Interested readers are referred to Appendix~\ref{sec:interleaving} for the detailed proof.
\end{remark}

It is known that for sufficiently large field size~$q$ (specifically $q = n^{\Theta(\min(k,n-k)L)}$), there always exists a linear $[n,k]$ code~$\code$ over $\mathbb{F}_q$ that is $\Int{L}$-MDS (and hence, $L$-MDS)~\cite{BGM23,BDG24}. This follows from the fact that the $\Int{L}$-MDS property translates into algebraic conditions on the $(n-k) \times n$ parity-check matrix $\bbH$ of $\code$: these conditions require a collection of square matrices to be nonsingular, and they hold for most matrices $\bbH$ if the underlying field is sufficiently large (in fact, they hold for most GRS codes). By Proposition~\ref{prop:lifting}, it follows that the interleaved code $\code^{\otimes \ell}$ is $\Int{L}$-MDS as well. This conclusion can be reached also from recalling that $\bbH$~is also a parity-check matrix of the code $\code^{\otimes \ell}$ when seen over $\Phi = \mathbb{F}_{q^{\ell}}$. Since the aforementioned algebraic conditions hold also over~$\Phi$, it follows that $\code^{\otimes \ell}$ is $\Int{L}$-MDS.

Next, we present our code construction.

\begin{construction}\label{Cons_L-MDS}
Let $N$, $T$ and $L$ be positive integers such that $N \geq 2T + 1$, $2 \leq L \leq N/T$ and $L\mid T$. Let~$\code$ be an $L$-MDS $[N,\rho N]$ code over a (sufficiently large) field $\mathbb{F}_q$, where
\begin{equation}\label{eq:k}
\rho =  1 - \Bigl( 1 + \frac{1}{L} \Bigr) \frac{T}{N}.
\end{equation}
Then the $\ell$-level interleaving $\code^{\otimes \ell}$ is an $L$-MDS code and is $(\tau,L)$-list decodable, where
\begin{equation}\label{eq:decodingradius}
\tau = \frac{L(N-\rho N)}{L+1}
\stackrel{\textrm{(\ref{eq:k})}}{=}
 T .
\end{equation}

Let $\bbG_1, \bbG_2, \ldots, \bbG_N$ be generator matrices of~$\code$, where each $\bbG_i$ contains the identity matrix in columns~$i$ through~$i+\rho N-1$ (with indices taken modulo~$N$ if they exceed~$N$). Since $\code$ is MDS by Theorem 3 in \cite{Roth22}, such generator matrices indeed exist. Letting $\Sigma = \mathbb{F}_q$ and $k = \rho N^2$, we write each information vector $\bbx \in \Sigma^{\rho N^2}$ as a concatenation of $N$ row vectors $\bbx_1, \bbx_2, \ldots, \bbx_N \in \Sigma^{\rho N}$ and define the encoder
$\bbf_N : \Sigma^{k} \rightarrow \Sigma^{N\times N}$ by
\begin{equation}
\label{eq:encoder}
\bbf_N(\bbx) =
\left(
\begin{array}{c}
\bbx_1 \cdot \bbG_1 \\
\bbx_2 \cdot \bbG_2 \\
\vdots      \\
\bbx_N \cdot \bbG_N
\end{array}
\right) .
\end{equation}
Clearly, $\bbf_N(\bbx)$ is a codeword in $\code^{\otimes N}$. 
\end{construction} 

Next, we show that the code $\code^{\otimes N}$ with the encoder $\bbf_N$ defined above is a $T$-VPEC code over $\mathbb{F}_q$ with parameters
\[
\left(N, \rho N^2, \frac{1}{\rho N}, \frac{LT}{N}\right).
\]
Recall that the rate--distortion pairs $(1/(N-T), F(T)/N)$ and $(1/(N-2T), 0)$ are achievable by Theorem~\ref{Fan's_main_result} and by $[N, N-2T]$ MDS codes, respectively. Thus, through a similar time-sharing argument as in \cite[Section II.B]{FKW17}, the rate--distortion pair $(R,D)$ that lies on the broken line connecting 
\[
\left( \frac{1}{N-T}, \frac{F(T)}{N} \right),\left(\frac{1}{N - (1 + L^{-1})T}, \frac{LT}{N}\right),~\text{and}~\left( \frac{1}{N-2T}, 0 \right) 
\]
can also be achieved. This confirms Corollary \ref{coro_Cons_L-MDS}.

\begin{IEEEproof}[Proof of Theorem \ref{Thm_Cons_L-MDS}]
For every information vector $\bbx \in \Sigma^k$ and every $j\in [N]$, take the $j$-th column of $\bbf_N(\bbx)$ as the $j$-th packet. That is,
\[
C_j = \begin{pmatrix}
      \bbx_1 \cdot \bbG_1(j),\,
      \bbx_2 \cdot \bbG_2(j),\,
      \ldots,\,
      \bbx_{N} \cdot \bbG_{N}(j)
    \end{pmatrix}^\top,
\]
where $\bbG_i(j)$ denotes the $j$-th column of $\bbG_i$. Recall that the code $\code^{\otimes N}$ is an $N\times N$ array code of dimension $k=\rho N^2$; thus, the rate of each packet equals to
\begin{equation}\label{eq:rateL}
    \frac{N}{\rho N^2}\stackrel{\textrm{(\ref{eq:k})}}{=}\frac{1}{N - (1 + L^{-1})T}.
\end{equation}

According to the definition of $\bbf_N$ in \eqref{eq:encoder}, when there are no packet errors, one can recover $\bbx_i$ by reading the $i$-th to $(i + \rho N - 1)$-st entries of the $i$-th row of the received word, and thus recover the original information vector $\bbx$. Thus, it remains to show that the distortion is at most $D$ in the case where there are at most $T$ packet errors.

Given $\bbY \in \Sigma^{N \times N}$ such that \[
d_{H}(\bbY, \bbf_N(\bbx)) \le T,
\]
a list decoder for $\code^{\otimes N}$ will generate at most~$L$ codewords $\Gamma \in \code^{\otimes N}$ such that $d_{H}(\bbY,\Gamma) \le T$. Then, by erasing from $\bbY$ all the columns in which there is disagreement among those codewords, at most $L T$ columns will be erased. The remaining columns are those where all the outputted codewords agree with $\bbY$, and they are the same as the correct codeword. Due to the specific structure of $\bbf_N(\bbx)$ in~(\ref{eq:encoder}), each column of $\bbf_N(\bbx)$ contains exactly $\rho N$ information symbols. Therefore, the total number of information symbols in the erased columns is at most
\[
\rho N\cdot LT = k\cdot \frac{LT}{N},
\]
which is at most $k$ since $L \leq N/T$. Hence, the distortion of the code in the case where there are at most $T$ packet errors is at most $\frac{LT}{N}$. This concludes the proof.
\end{IEEEproof}

\begin{remark}\label{rem:slack}
If we relax~(\ref{eq:k}) by requiring only that
\[
\rho \ge  1 - \Bigl( 1 + \frac{1}{L} \Bigr) \theta - \varepsilon ,
\]
for some (arbitrarily small yet) fixed small $\varepsilon > 0$, this would translate into subtracting~$\varepsilon$ from the denominator in the right-hand side of~\eqref{eq:rateL}. On the other hand, this allows us to select the alphabet $\Sigma$ to be a much smaller vector space over some $\mathbb{F}_q$ and take~$\code$ to be an $\mathbb{F}_q$-linear folded Reed--Solomon code over~$\Sigma$; see~\cite{Tamo24,CZ24}.
\end{remark}

%%%%%%%%%%%%%%%%%%%%%%%%%%%%%%%%%%%%%%%%%%%%%%

\subsection{VPEC codes with optimal rate--distortion trade-off}\label{sec: construction with opt trade-off}

Our second coding scheme can be regarded as a variant of the repetition code. Before delving into the formal description, we first introduce some additional notation and a toy example, which serves as a special case of the general coding scheme. %Throughout out this section, we assume that $k = (2T+1)m$, for some $m \in \Integers^+$.

We use $\ms{\cdot}$ to denote multisets. Let $B$ be a multiset of symbols in $\Sigma$ of size $b$, denoted by
$$B=\ms{x_1,x_2,\ldots,x_b}.$$
For each $y \in B$, we denote $F_{B}(y) \triangleq |\{i \in [b] : x_i = y\}|$ as the frequency of $y$ in $B$. Then, we use $MF_{B}$ to denote the most frequent elements in $B$.

\begin{example}\label{ex_1}
    Let $\Sigma$ be an arbitrary alphabet of size at least $2$ and let $T$ be a positive integer. For any message word $\bbx=(x_1,x_2,\ldots,x_{2T+1}) \in \Sigma^{2T+1}$ from the source, we encode it into the following $2T+1$ packets. For each $j \in [2T+1]$, the $j$-th packet is encoded as:
    \[
    C_j = (x_1, \ldots, x_{j-1}, x_{j+1}, \ldots, x_{2T+1})^\top,
    \]
    which is a word in $\Sigma^{2T}$.
    
    For decoding, we use the majority decoder to recover the value of $x_j$ for each $j\in [2T+1]$ and if it fails for some $j$, we then output $C_j$. Specifically, for each $j \in [2T+1]$, let 
    \begin{align*}
       \tilde{C}_j &= (\tilde{x}_{1,j}, \ldots, \tilde{x}_{j-1,j}, \tilde{x}_{j+1,j}, \ldots, \tilde{x}_{2T+1,j})^\top
    \end{align*}
    denote the received $j$-th packet, which may or may not be equal to $C_j$, where $\tilde{x}_{i,j} \in \Sigma$ for each $i \in [2T+1] \setminus \{j\}$. We view $\tilde{x}_{i,j}$ as the candidate of $x_i$ from the received $j$-th packet $\tilde{C}_{j}$. Note that for each $i\in [2T+1]$, there is only one packet $C_{i}$ which doesn't contain $x_i$. We denote $A_i=\ms{\tilde{x}_{i,j}:~j\in [2T+1]\setminus\{i\}}$ as the multiset of all the candidates of $x_i$. Then, the decoding algorithm is detailed in Algorithm \ref{alg:g0}.
    \begin{algorithm}
    \caption{Decoding Algorithm for codes in Example \ref{ex_1}}
    \label{alg:g0}
    \begin{algorithmic}
    \State \textbf{Input:} $(\tilde{C}_1, \ldots, \tilde{C}_{2T+1})$, parameter $T$
    \State \textbf{Output:} $\bby=(y_1, y_2, \ldots, y_{2T+1})$
    \State Initialize $I = \{1, 2, \ldots, 2T+1\}$

    \For{each $i \in I$}
    \State Let $v \gets MF_{A_i}$         (Most frequent element in $A_i$)
        \If{$F_{A_i}(v) \ge T + 1$}
            \State $y_i \gets v$
            \State $I \gets I \setminus \{i\}$
        \EndIf
        \EndFor

        \If{$|I| \le 0$}
        \State \Return $\bby=(y_1, \ldots, y_{2T+1})$
         \Else
        \State Pick any $i \in I$
        \For{each $j \in [2T+1] \setminus \{i\}$}
            \State $y_j \gets \tilde{x}_{j,i}$
        \EndFor
        \State $y_i \gets e$ (Assign erasure symbol)
        \EndIf
        \end{algorithmic}
    \end{algorithm}
\end{example}

For each message word $\bbx \in \Sigma^{2T+1}$, the encoder in Example \ref{ex_1} produces $2T+1$ packets $C_j \in \Sigma^{2T}$, which can be viewed as the columns in an $2T\times\left(2T+1\right)$ array over $\Sigma$. We claim that it achieves the rate--distortion pair $(2T/(2T+1),1/(2T+1))$, which is optimal with respect to the bound by Corollary \ref{coro1}. 

Clearly, the rate of each packet is $2T/(2T+1)$. Since there are at most $T$ altered packets, for each $i \in [2T+1]$, at least $T$ packets contain the correct $x_i$. Thus, by the decoding algorithm, if the most frequent element $MF_{A_i}$ in $A_i$ has frequency at least $T+1$, it must be correct. Otherwise, $MF_{A_i}$ shall have frequency $T$. Then, this implies that there are $T$ altered packets containing $x_i$. Thus, the received packet $\tilde{C}_i$, which doesn't contain $x_{i}$, must be correct. According to the decoding algorithm, we either recover all the correct $ x_i $ or obtain $ 2T $ correct message symbols $ x_i $ along with one erasure symbol, resulting in a distortion of at most $ \frac{1}{2T+1} $.

According to Algorithm~\ref{alg:g0}, determining the most frequent element is required for at most $N = 2T + 1$ sets $A_i$, each of size $|A_i| = N - 1 = 2T$. Using the Boyer-Moore majority vote algorithm~\cite{BM91}, which runs in $O(N)$ time per set, the overall time complexity of the decoding algorithm in Example~\ref{ex_1} is $O(N^2)$.

Next, we present the formal description of our general construction.

\begin{construction}\label{Cons2}
Let $\Sigma$ be an arbitrary alphabet of size at least $2$. Let $T$ and $s$ be positive integers such that $s \leq T$. For each $j\in [2T+1]$, we denote $S_j \triangleq \{ j, j+1, ..., j+s-1 \}$,
where we subtract $2T+1$ from an index if it exceeds $2T+1$, and $\overline{S}_j = [2T+1] \setminus S_j$.

For any message word $\bbx=(x_1,x_2,\ldots,x_{2T+1}) \in \Sigma^{2T+1}$ from the source, we encode it into the following $2T+1$ packets. For each $j \in [2T + 1]$, the $j$-th packet is encoded as the following word in $\Sigma^{2T + 1 - s}$:
\begin{align}\label{eq1_cons2}
    C_j & = (x_1,\ldots,x_{j-1},x_{j+s},\ldots,x_{2T+1})^{\top}=\left(\bbx|_{\overline{S}_j}\right)^{\top}.%\nonumber \\
    %&=\bbx|_{[N]\setminus [i,i+s-1]}.
\end{align}
For each $j\in [2T+1]$, we denote
\begin{align}\label{eq2_cons2}
   \tilde{C}_j&=(\tilde{x}_{1,j},\ldots,\tilde{x}_{j-1,j},\tilde{x}_{j+s,j},\ldots,\tilde{x}_{2T+1,j})^\top
\end{align}
as the received $j$-th packet, where $\tilde{x}_{i,j} \in \Sigma$ for each $i \in \overline{S}_j$. Note that for each $i \in [2T+1]$, there are exactly $s$ packets---namely, $C_{i}, C_{i-1}, \ldots, C_{i-s+1}$---that do not contain $x_i$. Therefore, for each $j \in \overline{S}_{i-s+1}$, the symbol $\tilde{x}_{i,j}$ can be viewed as a candidate for $x_i$ from the received packet $\tilde{C}_j$. Similar to Example~\ref{ex_1}, we select an appropriate symbol as $y_i$ from these candidates using a majority decoding approach. The full decoding procedure is detailed in Algorithm~\ref{alg:decoding Cons2}.
\end{construction}

\begin{algorithm}
\caption{Decoding Algorithm for codes by Construction \ref{Cons2}}
\label{alg:decoding Cons2}
\begin{algorithmic}%[1]
\State \textbf{Input:} $(\tilde{C}_1,\ldots,\tilde{C}_{2T+1})$, parameters $T$, $s$
\State \textbf{Output:} $\bby=(y_1, y_2, \ldots, y_{2T+1})$
\State Initialize $I = \{1, 2, \ldots, 2T+1\}$
\For{each $i \in I$}
    \State Let $v \gets MF_{A_i}$ (Most frequent element in $A_i$)
        \If{$F_{A_i}(v) \ge T + 1$}
            \State $y_i \gets v$
            \State $I \gets I \setminus \{i\}$
        \EndIf
    \EndFor

\While{$|I| > s$}
    \State Find $i_1, i_2 \in I$ such that:
    \Statex \hspace{\algorithmicindent} $i_2 - i_1 \in \{s, s+1, \ldots, 2T+1-s\} \mod 2T+1$
    \State Set $A_1=\ms{\tilde{x}_{i_1,j}:~j\in \overline{S}_{i_1-s+1}}$, $A_2=\ms{\tilde{x}_{i_2,j}:~j\in \overline{S}_{i_2-s+1}}$,
    $\xi_1 = MF_{A_1}$, $\xi_2 = MF_{A_1}$, $B_1 = \ms{\tilde{x}_{i_1,j} : j \in S_{i_2 - s + 1}}$, and $B_2 = \ms{\tilde{x}_{i_2,j} : j \in S_{i_1 - s + 1}}$

    \For{$\ell \in [2]$}
    \State Define Boolean expressions:
        \begin{align*}
        U_\ell &\triangleq \left[ F_{B_\ell}(\xi_\ell) \le T - F_{A_{3-\ell}}(\xi_{3-\ell}) \right] \\
        Z_\ell &\triangleq \left[ F_{B_\ell}(\xi_\ell) \ge s + F_{A_{3-\ell}}(\xi_{3-\ell}) - T \right]
        \end{align*}
        \If{$U_\ell$ is TRUE}
            \State $y_{i_{3 - \ell}} \gets MF_{B_{3 - \ell}}$
            \State $I \gets I \setminus \{ i_{3 - \ell} \}$
        \ElsIf{$Z_\ell$ is TRUE \textbf{or} $Z_{3-\ell}$ is FALSE}
                \State $y_{i_\ell} \gets \xi_\ell$
                \State $I \gets I \setminus \{ i_\ell \}$
        \EndIf
    \EndFor
\EndWhile

\For{$i \in I$}
    \State $y_i \gets e$ ~(Assign erasure symbol)
\EndFor

\end{algorithmic}
\end{algorithm}

\begin{remark}\label{rmk_k=m(2T+1)}
    Construction~\ref{Cons2} can be naturally extended to the case where $k = m(2T+1)$ for some $m \in \mathbb{Z}^{+}$. Specifically, we view $\mathbf{x} \in \Sigma^{m(2T+1)}$ as a concatenation of $2T+1$ words $
    \bbx_i = (x_{i,1}, x_{i,2}, \ldots, x_{i,m}) \in \Sigma^{m}$ for $i \in [2T+1]$, and perform $m$ independent transmissions. In the $\ell$-th transmission, send a batch of $2T+1$ packets encoding the message tuple $(x_{1,\ell}, x_{2,\ell}, \ldots, x_{2T+1,\ell})$. The decoder then runs the decoding algorithm separately on each batch of received words. This extension preserves both the per-packet rate and the distortion between the original message word $\mathbf{x} \in \Sigma^{m(2T+1)}$ and the final output.
\end{remark}

To show that Construction \ref{Cons2} yields $T$-VPEC codes with parameters as stated in Theorem \ref{thm_cons2_s=T}, we need the following lemmas.

\begin{lemma}\label{lem1_pre}
    Let $J$ be a subset of $\mathbb{Z}_{2T+1}$. If $|J|\geq s+1$ for some positive integer $s\leq T$, then there are $a,b\in J$ such that 
    $$a-b\in \{s,s+1,\ldots,2T+1-s \}\subseteq \mathbb{Z}_{2T+1}.$$
\end{lemma}

\begin{IEEEproof}
    We define a graph $G$ over $\mathbb{Z}_{2T+1}$ with edge set 
    $$E(G)\triangleq\left\{(a,b)\in (\mathbb{Z}_{2T+1})^{2}:~a-b\in \{ s,s+1,\ldots,2T+1-s \}\right\}.$$
    We claim that $G$ has independence number at most $s$. Then, for every vertex set $J$ of size at least $s+1$, there must be an edge among vertices in $J$.
    
    Clearly, each vertex $a\in \mathbb{Z}_{2T+1}$ has $2T+2-2s$ different neighbors $a+s,a+s+1,\ldots,a+2T+1-s$. Assume that $I$ is an independent set containing vertex $a$. Then, we know that
    \begin{align*}
        I\setminus\{a\}\subseteq\{a+1,\ldots,a+s-1\}\cup \{a+2T+2-s,\ldots,a+2T\}.
    \end{align*}
    Now, notice that for $i \in [s-1]$ we have $(a+2T+1-i)-(a+s-i)=2T+1-s$; that is, $(a+s-i,a+2T+1-i)$ is an edge in $G$. Thus, for each $1\leq i\leq s-1$, at most one of $\{a+s-i,a+2T+1-i\}$ is contained in $I$. Thus, we have $|I\setminus\{a\}|\leq s-1$ and this concludes the proof.
\end{IEEEproof}

\begin{lemma}\label{lem_cons2}
    Let $i_1, i_2 \in [2T+1]$ such that $i_2 - i_1 \in \{s, s+1, \ldots, 2T+1 - s\} \bmod{2T+1}$. Denote $A_1=\ms{\tilde{x}_{i_1,j}:~j\in \overline{S}_{i_1-s+1}}$, $A_2=\ms{\tilde{x}_{i_2,j}:~j\in \overline{S}_{i_2-s+1}}$,
    $\xi_1 = MF_{A_1}$, $\xi_2 = MF_{A_2}$, $B_1 = \ms{\tilde{x}_{i_1,j} : j \in S_{i_2 - s + 1}}$, and $B_2 = \ms{\tilde{x}_{i_2,j} : j \in S_{i_1 - s + 1}}$. For $\ell \in [2]$, define the Boolean expressions:
    \begin{align*}
        U_\ell &\triangleq \left[ F_{B_\ell}(\xi_\ell) \le T - F_{A_{3 - \ell}}(\xi_{3 - \ell}) \right], \\
        Z_\ell &\triangleq \left[ F_{B_\ell}(\xi_\ell) \ge s + F_{A_{3 - \ell}}(\xi_{3 - \ell}) - T \right].
    \end{align*}
    Then, the following statements hold for each $\ell \in [2]$:
    \begin{itemize}
        \item[1)] If $U_\ell = \text{TRUE}$, then $\xi_\ell$ is incorrect.
        \item[2)] If both $\xi_1$ and $\xi_2$ are incorrect, then $U_1 = U_2 = \text{TRUE}$.
        \item[3)] If $U_\ell = \text{FALSE}$ and $Z_\ell = \text{TRUE}$, then $\xi_\ell$ is correct.
        \item[4)] If $U_\ell = Z_{3 - \ell}= \text{FALSE}$, then $\xi_{\ell}$ is correct.
    \end{itemize}
\end{lemma}

\begin{IEEEproof}
First, we need the following observations. Note that for any $i_1,i_2\in [2T+1]$ such that $i_2-i_1\in \{s,s+1,\ldots,2T+1-s \} \mod{2T+1}$, we also have $i_1-i_2\in \{s,s+1,\ldots,2T+1-s \} \mod{2T+1}$. Thus, it holds that $S_{i_1-s+1}\cap S_{i_2-s+1}=\emptyset$ and
\begin{align*}
    S_{i_2-s+1}&\subseteq \overline{S}_{i_1-s+1},\\
    S_{i_1-s+1}&\subseteq \overline{S}_{i_2-s+1}.
\end{align*}
Recall that for each $i\in [2T+1]$, there are exactly $s$ packets $C_{i},C_{i-1},\ldots,C_{i-s+1}$ that do not contain $\bbx_i$. Therefore, the received packets $\tilde{C}_1,\tilde{C}_2,\ldots,\tilde{C}_{2T+1}$ have the following form:
\begin{equation}\label{eq2.1_lem_cons2}
    \begin{array}{ccccccc}
       \tilde{C}_{i_2-s+1} & =( & \cdots & \tilde{x}_{i_1,i_2-s+1} & \cdots & * & \cdots~)^\top, \\
       \vdots  &  &  & \vdots &  & \vdots & \\
       \tilde{C}_{i_2} & =( & \cdots & \tilde{x}_{i_1,i_2} & \cdots & * & \cdots~)^\top,\\
       \tilde{C}_{i_1-s+1}  & =( & \cdots & * & \cdots & \tilde{x}_{i_2,i_1-s+1} & \cdots~)^\top, \\
       \vdots  &  &  & \vdots &  & \vdots & \\
       \tilde{C}_{i_1} & =( & \cdots & * & \cdots & \tilde{x}_{i_2,i_1} & \cdots~)^\top,
    \end{array}
\end{equation}
where $*$ denotes the null symbol, and 
\begin{equation}\label{eq2.2_lem_cons2}
    \begin{array}{ccccccc}
    \tilde{C}_{j} & =( & \cdots & \tilde{x}_{i_1,j} & \cdots & \tilde{x}_{i_2,j} & \cdots~)^\top 
    \end{array}
\end{equation}
for the rest $j\in [2T+1]\setminus \left(S_{i_1-s+1}\cup S_{i_2-s+1}\right)$. 

Now, consider statement 1). Suppose that $U_1 = \text{TRUE}$. If $\xi_1$ were correct, that is, $\xi_1 = x_{i_1}$. Then, by the definition of $A_{\ell}$ and $B_{\ell}$, we have
\begin{align*}
    T+1 &\le |\{j \in [2T+1] :~ \tilde{C}_j=C_j \}| \\
        &\le F_{B_1}(x_{i_1}) + F_{A_2}(x_{i_2}) \\
        &= F_{B_1}(\xi_1) + F_{A_2}(x_{i_2}) \\
        &\le F_{B_1}(\xi_1) + F_{A_2}(\xi_2)
        \le T,
\end{align*}
which leads to a contradiction. Similarly, we have that $\xi_2$ is incorrect if $U_2 = \text{TRUE}$.

For statement 2), suppose both $\xi_1$ and $\xi_2$ are incorrect. Then, we have
\begin{align*}
    F_{B_1}(\xi_1) + F_{A_2}(\xi_2) &\le |\{j \in [2T+1] :~ \tilde{C}_j\neq C_j \}| \le T,\\
    F_{B_2}(\xi_2) + F_{A_1}(\xi_1) &\le |\{j \in [2T+1] :~ \tilde{C}_j\neq C_j \}| \le T,
\end{align*}
which directly implies $U_1 = U_2 = \text{TRUE}$. 

For statement 3), suppose that $U_1=\text{FALSE}$ and that $Z_1 = \text{TRUE}$. If $\xi_1$ were incorrect then $\xi_2$ had to be correct by statement 2). Among the packets that are sampled in $B_1$, there would be at most $s - F_{B_1}(\xi_1)$ correct packets. Hence, 
\begin{align*}
    T+1 &\le \left|\left\{j \in [2T+1] :~ \tilde{C}_j = C_j \right\}\right| \\
         &\le s - F_{B_1}(\xi_1) + F_{A_2}(\xi_2) \\
         &\le T,
\end{align*}
which leads to a contradiction, where the last inequality follows from $Z_1 = \text{TRUE}$. Similarly, we have that $\xi_2$ is correct if $U_2=\text{FALSE}$ and $Z_2 = \text{TRUE}$.

For statement 4), suppose that $U_1= Z_2= \text{FALSE}$. If $\xi_1$ were incorrect then $\xi_2$ had to be correct by statement 2). Among the packets that are sampled in $B_2$, there would be at least $s - F_{B_2}(\xi_2)$ incorrect
packets. Thus,
\begin{align*}
    T &\ge |\{j \in [2T+1] :~ \tilde{C}_j\neq C_j \}| \\
    &\ge s - F_{B_2}(\xi_2) + F_{A_1}(\xi_1)
     >   T ,
\end{align*}
which leads to a contradiction, where the last inequality follows from $Z_2= \text{FALSE}$. Similarly, we have that $\xi_2$ is correct if $U_2=Z_1=\text{FALSE}$.

This completes the proof.
\end{IEEEproof}

Now, we proceed with the proof of Theorem \ref{thm_cons2_s=T}

\begin{IEEEproof}[Proof of Theorem \ref{thm_cons2_s=T}]
By Remark~\ref{rmk_k=m(2T+1)}, it suffices to show that Construction \ref{Cons2} yields $T$-VPEC codes that achieve the rate--distortion pair $\left(1-s/(2T+1), s/(2T+1)\right)$.

For each message word $\bbx \in \Sigma^{2T+1}$, the encoder in Construction \ref{Cons2} produces $2T+1$ packets $C_j \in \Sigma^{2T+1-s}$. Clearly, the rate of each packet $C_j$ is $1-s/(2T+1)$. Next, we show that the output $\bby \in \tilde{\Sigma}^{2T+1}$ of the Algorithm \ref{alg:decoding Cons2} is the correct message word $\bbx$ when there are no errors, and the erasure distortion between $\bby$ and $\bbx$ satisfies $\Delta(\bbx,\bby)\leq s/(2T+1)$ when there are at most $T$ erroneous packets.

First, in each iteration of the ``while'' loop of Algorithm~\ref{alg:decoding Cons2}, we note the following two facts:
\begin{itemize}
    \item By Lemma~\ref{lem1_pre}, one can always find $i_1, i_2 \in I$ such that $i_2 - i_1 \in \{s, s+1, \ldots, 2T+1-s\} \bmod 2T+1$.
    \item In the inner ``for'' loop, the ``if--else if'' condition will be satisfied for at least one $\ell\in \{1,2\}$.
\end{itemize}
Thus, the ``while'' loop in Algorithm~\ref{alg:decoding Cons2} can always proceed as long as $|I| > s$, and the value of $|I|$ decreases by at least one in each iteration. This ensures that Algorithm~\ref{alg:decoding Cons2} always terminates.

Now, consider the case where there are no errors; that is, $\tilde{C}_j = C_j$ for every $j \in [2T+1]$. Then, it holds that $x_i = MF_{A_i}$ and $F_{A_i}(x_i) = |A_i| = 2T+1 - s$, for each $i \in [2T+1]$. Since $s \leq T$, we have $F_{A_i}(x_i) \geq T+1$ for every $i \in [2T+1]$. Therefore, we would have $|I| = 0$ after completing the first ``for'' loop. The algorithm then terminates and outputs $\mathbf{y} = (x_1, x_2, \ldots, x_{2T+1})$, which is the correct message word.

Next, we consider the case where there are at most $T$ erroneous packets.

Fix an index $i_1 \in [2T+1]$ such that $y_{i_1} \neq e$. The proof consists of the following two parts:
\begin{itemize}
    \item Suppose $y_{i_1}$ is output in the first ``for'' loop of Algorithm~\ref{alg:decoding Cons2}. Then we have $F_{A_{i_1}}(y_{i_1}) \geq T+1$. Since there are at most $T$ erroneous packets, this implies $y_{i_1} = x_{i_1}$.
    
    \item Suppose $y_{i_1}$ is output in the inner ``for'' loop of the ``while'' loop. Then there exists some $i_2 \in [2T+1]$ such that $i_2 - i_1 \in \{s, s+1, \ldots, 2T+1 - s\} \mod 2T+1$. Furthermore, one of the following three conditions must hold: 1) $U_2 = \text{TRUE}$, 2) $U_1 = \text{FALSE}$ and $Z_1 = \text{TRUE}$, 3) $U_1 = Z_2 = \text{FALSE}$. In cases 2) and 3), the correctness of $y_{i_1} = \xi_1$ follows directly from the third and fourth statements of Lemma~\ref{lem_cons2}, respectively. For case 1), by the first statement of Lemma~\ref{lem_cons2}, $\xi_2$ is incorrect. This implies that at least $(2T+1 - s)/2$ members of $A_2$ are incorrect. Consequently, at most $(s - 1)/2$ members of $B_1$ can be incorrect, which in turn implies that $MF_{B_1}$ is correct.
\end{itemize}
Therefore, each $y_i \neq e$ in the output is correct. Since there are at least $2T+1 - s$ non-erasure symbols in the output, the erasure distortion between the output $\bby$ and $\bbx$ satisfies $\Delta(\bbx, \bby) \leq s/(2T+1)$.

Finally, we conclude the proof by analyzing the time complexity of Algorithm~\ref{alg:decoding Cons2}. 

According to its description, in the first ``for'' loop, one needs to determine the most frequent element of $A_i$ for all $i \in [2T+1]$. In the inner ``for'' loop of the ``while'' loop, one needs to determine the most frequent element for at most $|I| (2T + 2 - 2s)$ different set pairs $(B_1, B_2)$. Moreover, finding $i_1, i_2 \in I$ such that $i_2 - i_1 \in \{s, s+1, \ldots, 2T+1 - s\} \mod 2T+1$ requires $O(|I|^2)$ time, and computing the values of $U_1, U_2, Z_1, Z_2$ for each such pair $(i_1, i_2)$ incurs an additional constant computational cost.

Note that each set $A_i$ has size $2T+1 - s$, and for each pair $(i_1, i_2)$ satisfying the condition, both $B_1$ and $B_2$ have size $s$. Hence, by the well-known Boyer-Moore majority vote algorithm~\cite{BM91}, which operates with a time complexity linear in the size of the set, the overall time complexity of the decoding algorithm is at most
\[
(2T+1) \cdot O(2T+1 - s) + |I| (2T + 2 - 2s) \cdot O(s) + O(T^2) = O(T^3).
\]
This completes the proof.
\end{IEEEproof}

\section{Conclusion and further research}\label{sec: conclusion}

The study presented in this paper is inspired by \cite{AW17} and \cite{FKW17}, which explore achievable rate--distortion pairs and the construction of corresponding VPEC coding schemes. Hence, our primary focus is on investigating the rate--distortion trade-off and constructing VPEC codes with good rate--distortion performance. %Although some progresses has been made in this topic, many questions remain open. 
In the following, we highlight several directions for future research.
\begin{enumerate}
    \item[1.] The primary open question remains the improvement of bounds on the rate--distortion trade-off for the $T$-VPEC problem. Although we have shown that the bound in Theorem~\ref{main_Thm1} is tight for the case $N = 2T + 1$ via Construction~\ref{Cons2}, there remains a gap between this bound and the rate--distortion pairs achieved by Construction~\ref{Cons_L-MDS} for general parameter regimes (see Figure~\ref{figure2} for an illustration).
    \item[2.] In Construction \ref{Cons2} (or Remark \ref{rmk_k=m(2T+1)}), a message word is equally partitioned into $2T+1$ fragments, every such fragment is contained in exactly $2T+1-s$ packets, and every packet consists of exactly $2T+1-s$ message fragments. Such kinds of combinatorial structures are commonly seen in combinatorial design theory. Thus, a possible direction is to employ similar combinatorial structures from this field to design VPEC codes with better rate--distortion performance.
    \item[3.] In the original setting of the variable packet-error code problem in \cite{AW17}, the rates of different packets can be different. This fits the communication scenario where different paths in the network have different channel environments. Thus, it's also interesting to investigate the achievable rate--distortion vectors and constructions of corresponding coding schemes in this general setting.
    %\item[3.] In Corollary \ref{coro1} (or Theorem \ref{main_Thm1}), our bound on the rate--distortion trade-off contains two parts, $\left(1-\frac{1}{k}\right)\frac{1}{n-2T}$, which comes from the partial decodability, and $\frac{1}{n-T}$, which comes from the $T$-error-detectability. These two parts are proved separately and both rely on the characterization of the minimum distance of the code. Is it possible to leverage Lemma \ref{lem_VPEC-code_property1}, which provides a unified characterization of the code satisfying these two properties, and further improve the bound by Corollary \ref{coro1}?
\end{enumerate}

\section*{Acknowledgment}

We extend our gratitude to Prof. Aaron B. Wagner for introducing us to the question of variable packet-error coding.

\appendices

\section{The Diametric Theorem and the proof of Corollary \ref{coro1}}\label{sec: Diametric Thm and proof of Coro1}

\begin{theorem}[{\cite[Diametric Theorem]{AK98}}]\label{the_diametric_theorem}
    For positive integers $ n, d $, and $ q \geq 2 $ satisfying $ n-d \geq 1 $, let $ r \geq 0 $ be the largest integer such that 
    \begin{equation}\label{eq1_Diametric_thm}
        2r \leq \min\left\{d, \frac{2(n-d)-q}{q-2}\right\},
    \end{equation}
    where we set $ \frac{1}{q-2} = \infty $ for $ q = 2 $. Denote $ \mathscr{K}_{r} $ as the following set of words in $ [q]^n $:
    \begin{equation}\label{eq2_Diametric_thm}
        \mathscr{K}_{r} = \{\bbx \in [q]^n : \abs{\{i \in [n-d+2r] : x_i = 1\}} \geq n-d+r\}.
    \end{equation}
    Then, $ Ant_{q}(n,d) = |\mathscr{K}_{r}| $.
\end{theorem}

\begin{IEEEproof}[Proof of Corollary \ref{coro1}]
    Note that in Theorem \ref{the_diametric_theorem}, when $q \geq \frac{2}{3}(n-d)+2$, we have
    \begin{align*}
        \frac{2(n-d)-q}{q-2}&=\frac{2(n-d)-2}{q-2}-1\\
        &\leq \frac{2(n-d)-2}{\frac{2}{3}(n-d)}-1\\
        &<2.
    \end{align*}
    Thus, in this case, the largest integer $r \geq 0$ satisfying \eqref{eq1_Diametric_thm} is $r = 0$. This implies that $Ant_{q}(n,d) = |\mathscr{K}_{0}|$ when $q \geq \frac{2}{3}(n-d)+2$. Moreover, from \eqref{eq2_Diametric_thm}, we know that $\mathscr{K}_{0} = \{\bbx \in [q]^n : x_i = 1 \text{ for all } i \in [n-d]\}$. Consequently, the result follows directly from $|\mathscr{K}_{0}| = q^d$ and the first two parts of Theorem \ref{main_Thm1}.
    % When $q=2$, the largest integer $ r \geq 0 $ satisfying \eqref{eq1_Diametric_thm} is $ r = \lfloor\frac{d}{2}\rfloor $. Then, by Theorem \ref{the_diametric_theorem}, we have 
    % \begin{align*}
    %     Ant_{2}(k,kD) &= |\mathscr{K}_{\lfloor\frac{kD}{2}\rfloor}|\leq \sum_{i=0}^{\lceil\frac{kD}{2}\rceil}{k\choose i} \\
    %     & \leq2^{kH_2\left(\frac{D}{2}\right)}.
    % \end{align*}
    % Then, by Theorem \ref{main_Thm1}, this leads to 3).
\end{IEEEproof}

\section{Asymptotic behaviors of $(R^{O},D)$}\label{sec: asympto}

By Theorem~\ref{Fan's_main_result}, when $T/N = \theta$, the overall rate--distortion pair $(R^{O}, D)$ achieved by polytope codes becomes
\begin{equation}\label{eq_asympto1}
    \left(R^{O}, \frac{(1 - 2\theta)^{-1} - R^{O}}{(1 - 2\theta)^{-1} - (1 - \theta)^{-1}} \cdot \frac{F(T)}{N} \right),
\end{equation}
for $\frac{1}{1 - \theta} \leq R^{O} \leq \frac{1}{1 - 2\theta}$. Since $F(T) = T + \left\lfloor \frac{T^2}{4} \right\rfloor + 1$, we have $F(T)/N \to \infty$ as $N \to \infty$, causing the distortion in \eqref{eq_asympto1} to also diverge. Meanwhile, the overall rate--distortion pair achieved by MDS codes is
\begin{equation}\label{eq_asympto2}
    \left(R^{O}, \frac{1 - \theta}{\theta} \left(1 - (1 - 2\theta) R^{O} \right) \right),
\end{equation}
which is clearly better than \eqref{eq_asympto1}.

Moreover, by Theorem~\ref{Thm_Cons_L-MDS}, the following overall rate--distortion pair is achievable by $L$-MDS codes:
\[
\left( \frac{L}{L - (L + 1)\theta}, \, L\theta \right).
\]
When $R^O = \frac{L}{L - (L + 1)\theta}$, the distortion in~\eqref{eq_asympto2} becomes
\begin{align*}
    \frac{1 - \theta}{\theta} \left(1 - \frac{1 - 2\theta}{1 - (1 + \frac{1}{L})\theta} \right) 
    &= \frac{1 - \theta}{\theta} \cdot \frac{(1 - \frac{1}{L})\theta}{1 - (1 + \frac{1}{L})\theta} \\
    &= \left(1 - \frac{1}{L} \right) \cdot \frac{1 - \theta}{1 - (1 + \frac{1}{L})\theta} \\
    &= \left(1 - \frac{1}{L} \right) \cdot \left(\frac{L}{L+1} + \frac{1/(L+1)}{1 - (1 + \frac{1}{L})\theta} \right) \\
    &= \frac{L - 1}{L + 1} + \frac{1}{L + 1} \cdot \frac{L - 1}{L - (L + 1)\theta}. %\label{eq_asympto3}
\end{align*}

\begin{figure}[htbp]
    \centering
    \subfigure{
    \includegraphics[width=0.44\textwidth]{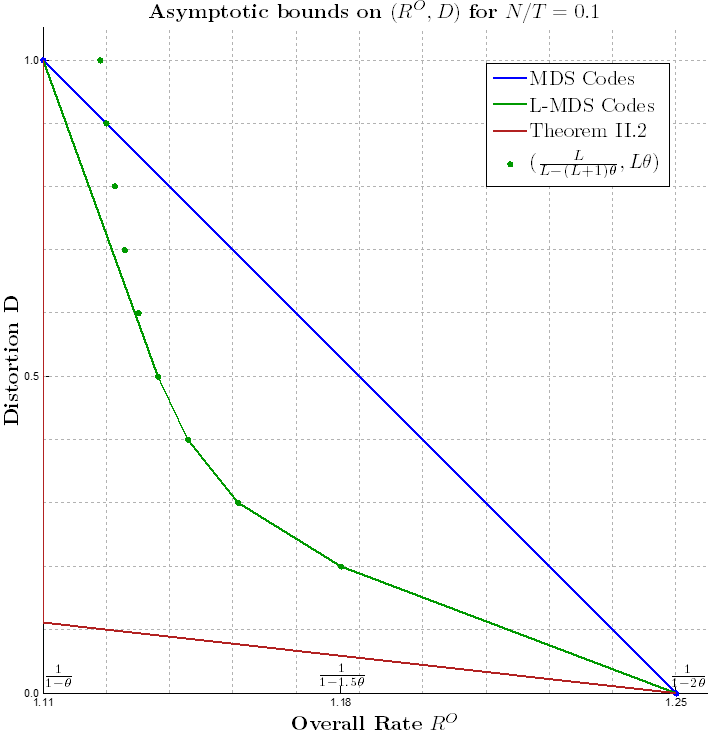}
    }    
    \subfigure{
    \includegraphics[width=0.44\textwidth]{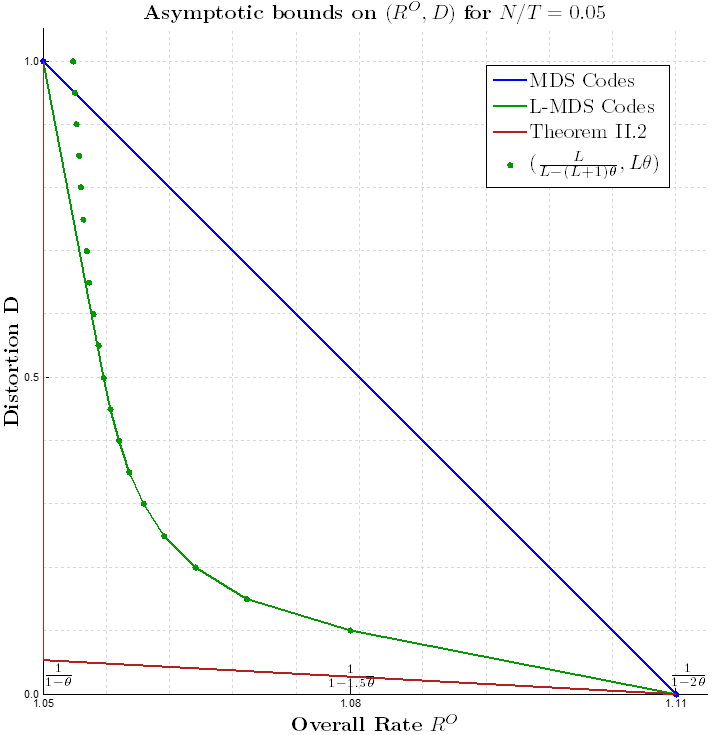}
    }
    \caption{Asymptotic bounds on the overall rate--distortion trade-off from Theorem~\ref{main_Thm1}, MDS codes and $L$-MDS codes (Corollary~\ref{coro_Cons_L-MDS}), for $T/N = 0.1$, and $T/N = 0.05$.}
    \label{figure3}
\end{figure}

Consider the function
\[
f(x) = Lx - \frac{L - 1}{L + 1} - \frac{1}{L + 1} \cdot \frac{L - 1}{L - (L + 1)x},
\]
we have
\[
f'(x) = L - \frac{(L - 1)(L + 1)^2}{\left(L - (L + 1)x\right)^2},
\]
which is at least $L - 1$ for $L \geq 1$ and $x \leq 1/(L + 1)$. That is, for $L \geq 2$, the function $f(x)$ is increasing in $x$ over the interval $x \leq 1/(L + 1)$. Therefore, we obtain
\[
f(\theta) \leq f\left(\frac{1}{L + 1}\right) = 0,
\]
whenever $\theta \leq 1/(L + 1)$. This implies that $L$-MDS codes achieve a better overall rate--distortion trade-off than MDS codes when $\theta \leq 1/(L + 1)$. For illustrative comparisons, see Figure~\ref{figure3}.

\section{Preserving strong list decodability under interleaving}
\label{sec:interleaving}

Let $\ell$, $n$, $d$ be given positive integers
such that $d \le n$ and let $\tau \in \Integers/3$ be positive such that
\begin{equation}
\label{eq:2/3}
\tau \le \frac{2d-1}{3} .
\end{equation}
For $i \in \Int{\ell}$, let $\Sigma_i$ be a finite alphabet
(which, without loss of generality, is assumed to be
an Abelian group)
and let $\code_i$ be a code in $\Sigma_i^n$ of
minimum distance${} \ge d$
that is strongly-$(\tau,2)$-list decodable.

%The following proposition extends the result of Proposition~\ref{prop:lifting} to the setting of strong list decodability.

\begin{proposition}
\label{prop:interleaving}
The interleaved code $\otimes_{i \in \Int{\ell}} \code_i$ is strongly-$(\tau,2)$-list decodable over $\otimes_{i \in \Int{\ell}} \Sigma_i$.
\end{proposition}

\begin{IEEEproof}
It suffices to prove the claim for $\ell = 2$, as the general case follows by induction.

Given a $2 \times n$ array
\[
\bbY =\left(
\begin{array}{c}
\bby_1 \\ \bby_2
\end{array}
\right) ,
\]
with rows $\bby_1 \in \Sigma_1^n$ and $\bby_2 \in \Sigma_2^n$, we show that there can be at most two code arrays $\Gamma \in \code_1 \otimes \code_2$ such that $d_{H}(\bbY,\Gamma) \le \tau$.

Suppose to the contrary that there exist three distinct code arrays $\Gamma_1, \Gamma_2, \Gamma_3 \in \code_1 \otimes \code_2$ such that
\begin{equation}\label{eq:assumption1}
\sum_{r \in \Int{3}}
d_{H}(\bbY, \Gamma_r) \le 3 \tau .
\end{equation}
For $r \in \Int{3}$, write
\[
\Gamma_r =
\left(
\begin{array}{c}
\bbc_1^{(r)} \\ \bbc_2^{(r)}
\end{array}
\right) ,
\]
where $\bbc_i^{(r)} \in \code_i$, for $i \in \Int{2}$; also, let
\[
\cA_i^{(r)} = \Support \bigl( \bby_i - \bbc_i^{(r)} \bigr) .
\]
Then (\ref{eq:assumption1}) is equivalent to
\begin{equation}\label{eq:assumption2}
\sum_{r \in \Int{3}}
\bigl| \cA_1^{(r)} \cup \cA_2^{(r)} \bigr| \le 3 \tau .
\end{equation}
Yet, since $\code_1$ is strongly-$(\tau,2)$-list decodable, at least two among the codewords $\bbc_1^{(1)}, \bbc_1^{(2)}, \bbc_1^{(3)}$ must be the same. Without loss of generality assume that $\bbc_1^{(2)} = \bbc_1^{(3)}$, which means that
\begin{equation}\label{eq:identical1}
\cA_1^{(2)} =  \cA_1^{(3)}
\end{equation}
and that $\bbc_2^{(2)} \ne \bbc_2^{(3)}$. Similarly, at least two among the codewords $\bbc_2^{(1)}, \bbc_2^{(2)}, \bbc_2^{(3)}$ must be the same, say $\bbc_2^{(1)} = \bbc_2^{(2)}$, which means that
\begin{equation}\label{eq:identical2}
\cA_2^{(1)} =  \cA_2^{(2)}
\end{equation}
and that $\bbc_1^{(1)} \ne \bbc_1^{(2)}$.

Write
\[
\begin{array}{rclcl}
\cV & =  & \cA_1^{(1)} , && \\
\cX & =  & \cA_1^{(2)} \setminus \cV
& \stackrel{\textrm{(\ref{eq:identical1})}}{=} &
\cA_1^{(3)} \setminus \cV , \\
\cY & =  & \cA_2^{(3)} , && \\
\cZ & =  & \cA_2^{(1)} \setminus \cY
& \stackrel{\textrm{(\ref{eq:identical2})}}{=} &
\cA_2^{(2)} \setminus \cY .
\end{array}
\]
From~(\ref{eq:assumption2}) we get:
\[
\bigl| \cV \cup \cZ \bigr|
+ \bigl| \cX \cup \cZ \bigr|
+ \bigl| \cY \bigr|
\le 3 \tau ,
\]
i.e.,
\[
\bigl| \cZ \setminus \cV \bigr| + \bigl| \cV \bigr|
+ \bigl| \cZ \setminus \cX \bigr| + \bigl| \cX \bigr|
+ \bigl| \cY \bigr|
\le 3 \tau .
\]
Since $\cX \cap \cV = \emptyset$, we have
\[
\bigl| \cZ \bigr|
\le \bigl| \cZ \setminus \cX \bigr| + \bigl| \cZ \setminus \cV \bigr|
\]
and, so, the last two inequalities result in
\[
\bigl| \cV \bigr| + \bigl| \cZ \bigr|
+ \bigl| \cX \bigr| + \bigl| \cY \bigr|
\le 3 \tau .
\]
On the other hand, since $\code_1$ and~$\code_2$ have minimum distance${} \ge d$, we also have
\[
\bigl| \cV \bigr| + \bigl| \cX \bigr|
= \bigl| \cA_1^{(1)} \cup \cA_1^{(2)} \bigr|
\ge
\bigl| \Support \bigl(\bbc_1^{(1)} - \bbc_1^{(2)} \bigr) \bigr| \ge d
\]
and
\[
\bigl| \cY \bigr| + \bigl| \cZ \bigr|
= \bigl| \cA_2^{(3)} \cup \cA_2^{(2)} \bigr|
\ge
\bigl| \Support \bigl(\bbc_2^{(3)} - \bbc_2^{(2)} \bigr) \bigr| \ge d.
\]
Combining the last three inequalities leads to
\[
2d \le
\bigl| \cV \bigr| + \bigl| \cX \bigr|
+ \bigl| \cY \bigr| + \bigl| \cZ \bigr|
\le 3 \tau
\stackrel{\textrm{(\ref{eq:2/3})}}{\le}
2d-1 ,
\]
which is a contradiction.
\end{IEEEproof}

\bibliographystyle{IEEEtran}
\bibliography{biblio}

\end{document}